\documentclass[reprint, amsmath, amssymb, aps, nofootinbib, twocolumn,superscriptaddress]{aastex631}

\usepackage{graphicx}% Include figure files
\usepackage{dcolumn}% Align table columns on decimal point
\usepackage{bm}% bold math
\usepackage{comment}
\usepackage{aas_macros}
\usepackage{tabularx}
\usepackage{multirow}
\usepackage{xcolor}
\usepackage{fontawesome5}

\newcolumntype{P}[1]{>{\centering\arraybackslash}p{#1}}

\definecolor{orcidlogocol}{rgb}{0.65, 0.807, 0.223}
\newcommand{\orcid}[1]{$\,$\href{https://orcid.org/#1}{\textcolor{orcidlogocol}{\faOrcid}}}

\begin{document}

\title{Learning cosmology and clustering with cosmic graphs}

\author{Pablo Villanueva-Domingo\orcid{0000-0002-0936-4279}}
\email{pablo.villanueva.domingo@gmail.com}
\affiliation{Computer Vision Center - Universitat Aut\`onoma de Barcelona (UAB), 08193 Bellaterra, Barcelona, Spain}

\author{Francisco Villaescusa-Navarro\orcid{0000-0002-4816-0455}}
\email{fvillaescusa@flatironinstitute.org}
\affiliation{Center for Computational Astrophysics, Flatiron Institute, 162 5th Avenue, New York, NY, 10010, USA}
\affiliation{Department of Astrophysical Sciences, Princeton University, Peyton Hall, Princeton NJ 08544, USA}

\begin{abstract}
We train deep learning models on thousands of galaxy catalogues from the state-of-the-art hydrodynamic simulations of the CAMELS project to perform regression and inference. We employ Graph Neural Networks (GNNs), architectures designed to work with irregular and sparse data, like the distribution of galaxies in the Universe. We first show that GNNs can learn to compute the power spectrum of galaxy catalogues with a few percent accuracy. We then train GNNs to perform likelihood-free inference at the galaxy-field level. Our models are able to infer the value of $\Omega_{\rm m}$ with a $\sim12\%-13\%$ accuracy just from the positions of $\sim1000$ galaxies in a volume of $(25~h^{-1}{\rm Mpc})^3$ at $z=0$ while accounting for astrophysical uncertainties as modelled in CAMELS. Incorporating information from galaxy properties, such as stellar mass, stellar metallicity, and stellar radius, increases the accuracy to $4\%-8\%$. Our models are built to be translational and rotational invariant, and they can extract information from any scale larger than the minimum distance between two galaxies. However, our models are not completely robust: testing on simulations run with a different subgrid physics than the ones used for training does not yield as accurate results.
\end{abstract}

\section{Introduction}

The standard model of cosmology is the current most accepted theoretical framework that accurately describes a very large set of cosmological observations. This model contains a set of free parameters characterizing fundamental properties on the Universe, such as its geometry, expansion rate, and composition. Constraining the value of these parameters is a major priority in cosmology; we expect to improve our knowledge on fundamental physics by measuring these parameters with the highest accuracy possible. 

The spatial distribution of galaxies in the Universe contains a wealth of information about the value of the cosmological parameters. That information is typically extracted using summary statistics like the power spectrum, that are known to be suboptimal descriptors of the statistical properties of non-Gaussian density fields, like the galaxy density field on non-linear scales. Recently, a large effort has been carried out to compare how much cosmological information can be extracted from different summary statistics of the web, such as correlation functions and higher momenta \citep{Villaescusa_Navarro_2020, Samushia_2021, Gualdi_2020, Gualdi_2021, Hahn_2020, Hahn_2021}, counts in cells \citep{Banerjee_2020, Banerjee_2021, Uhlemann_2020}, Minkowski functionals \citep{vicinanaza_2019, Liu_2022, Marques_2019}, peak counts \citep{Zack_2019, Ajani_2020, Harnois_2021a, Harnois_2021}, wavelets \citep{Valgiannis_2021, 2022arXiv220407646E, Sihao_2021}, via the kinetic Sunyaev Zel’dovich effect \citep{Kuruvilla_2021, Kuruvilla_2021b, Giri_2020} or through other approaches \citep{Banerjee_2019, Bayer_2021, Allys_2020, Naidoo_2021, Bella_2020, massara_2021, Liu_2019, Lee_2020a, Lee_2020, Gemma_2019,  simpson_2011, simpson_2013, Friedrich_2020, Dai_2020}.

Alternatively, machine learning (ML) methods can be used to optimally extract information from the field itself without using summary statistics. This task has been carried out, typically using Convolutional Neural Networks (CNNs), for density fields \citep{Siamak_16, 2021arXiv210909747V, 2021arXiv210910360V, Lazanu_2021, Hortua_2021, 2021JCAP...11..049M}, weak lensing maps \citep{Schmelzle_17, Gupta_18, Ribli_19, Fluri_19, Jose_2020, Niall_2020, Lu_2022}, cosmic microwave background maps \citep{he2018analysis}, and 21 cm maps \citep{2019MNRAS.484..282G, Sultan_2019, 2021ApJ...907...44V} among others.

In this work we use machine learning to extract the maximum information from galaxy catalogues without using summary statistics. For this task we employ Graph Neural Networks (GNNs), that are capable of dealing with irregular and sparse data \citep{2021arXiv210413478B, 2018arXiv180601261B, HamiltonBook}, like the spatial distribution of galaxies. The models we use here employ the so-called message passing scheme, where information from a defined neighborhood is shared with a particular galaxy, exploiting global and local relations. While these models are already by construction permutation invariant, we build them to respect the physical symmetries associated to this task: rotational and translational invariance. Another advantage of GNNs is that they do not impose any cut in scale. With CNNs, the grid size determines the minimum scale\footnote{While it is possible to take a finer grid, making the cutoff scale smaller, it comes at the cost of using more memory and increasing the sparsity of the data.} where information can be extracted, while there is not such scale cutoff in GNNs, where galaxies can be arbitrarily close and the model can extract information from features on such scales. Due to their multiple advantages, GNNs are increasingly being applied in cosmology and astrophysics \citep{2019arXiv190905862C, Cranmer:2020wew, Beck2019RefinedRR, Cranmer:2021pve, GNN_CAMELS, GNN_MW_M31}.

Previous works have already tried to infer cosmological parameters from 3D galaxy distributions using machine learning. For instance, CNNs have been applied to galaxy catalogues created combining Halo Occupation Distribution (HOD) models with N-body simulations \citep{Ntampaka_19}. However, we believe that our approach represents a step forward for several reasons. First, we use a deep learning architecture based on GNNs that we believe is more appropriate to deal with the sparse and irregular data associated to galaxies. Second, our galaxy catalogues come from state-of-the-art hydrodynamic simulations of the CAMELS project \citep{villaescusanavarro2020camels} that model galaxy positions and properties more accurately than other methods. Third, we use GNNs to extract information not only from galaxy clustering, but also from galaxy properties such as stellar mass, stellar metallicity, and stellar radius. Fourth, we construct our model to be invariant under the desired symmetries: translations, rotations, and permutations. Fifth, we perform likelihood-free inference to obtain a confident estimate of the expected uncertainty of our models. Sixth, we perform robustness tests to establish whether the model can potentially be used with real data or whether it is too dependent on the type of data used for training.

This paper is organized as follows. In Sec. \ref{sec:data} we describe the data we use in this work. The architecture of our model is described in Sec. \ref{sec:model}. We first show in Sec. \ref{sec:ps} that graph neural networks can learn estimators of galaxy clustering like the power spectrum. We then show the results of inferring cosmological parameters from galaxy distributions in Sec. \ref{sec:cosmology_results}. Finally, we conclude and discuss the main results of this work in \ref{sec:discussion}.

%%%%%%%%%%%%%%%%%%%%%%%%%%%%%%%%%%%%%%%%%%%%%%%%%%%%%%%
%%%%%%%%%%%%%%%%%%%%%%%%%%%%%%%%%%%%%%%%%%%%%%%%%%%%%%%
\section{Data}
\label{sec:data}

\begin{figure*}[th!]
\begin{center}
\includegraphics[width=0.49\linewidth]{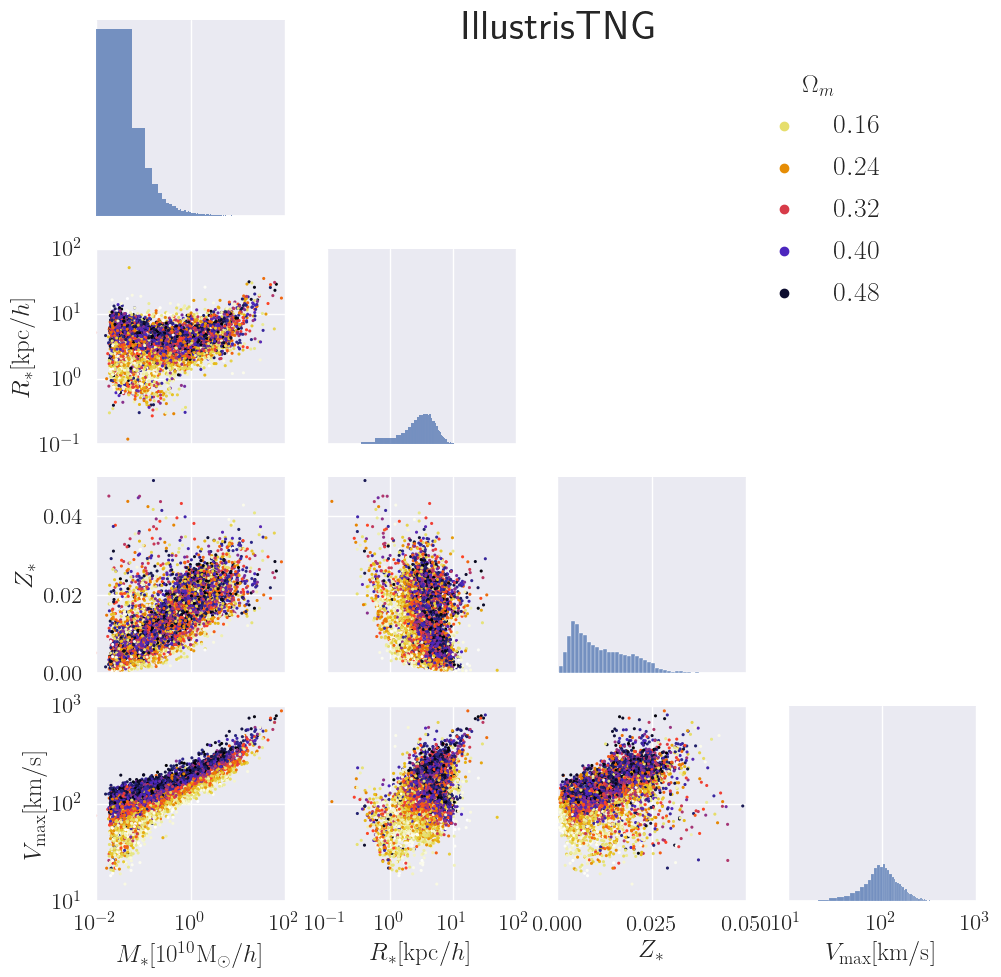}
\includegraphics[width=0.49\linewidth]{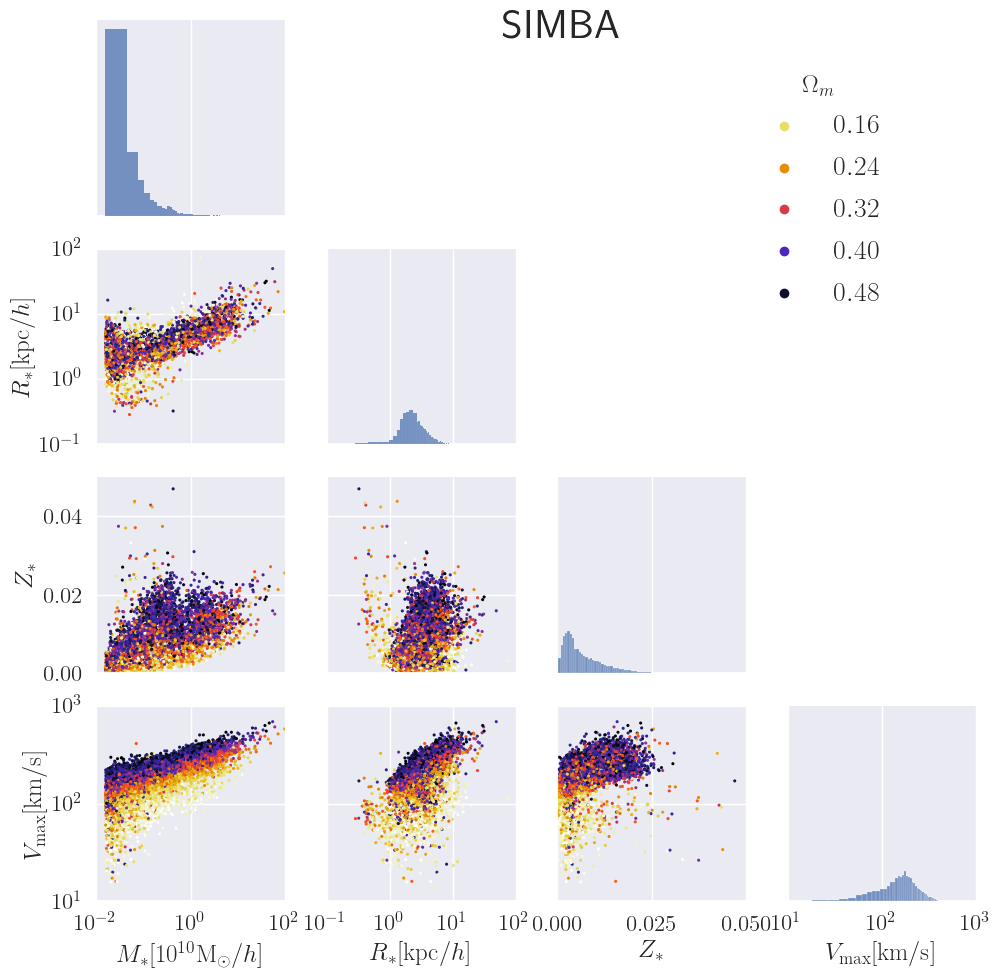}
\caption{This figure shows 1D and 2D distributions of galaxy properties in the IllustrisTNG (left) and SIMBA (right) simulations. Each point represents a galaxy, and it is color-coded according to the value of $\Omega_{\rm m}$. The considered galaxy properties are: stellar mass, $M_*$, stellar half-mass radius, $R_{*}$, stellar metallicity, $Z_*$, and maximum circular velocity, $V_{\rm max}$. As can be seen, in the CAMELS simulations, galaxy properties occupy different regions in the feature space depending on the value of $\Omega_{\rm m}$. Hence, galaxy properties may be used to used to infer the value of the cosmological parameters.}
\label{fig:jointplot}
\end{center}
\end{figure*}

In this work we have used galaxy catalogues from simulations of the CAMELS project \citep{villaescusanavarro2020camels,2022arXiv220101300V}. Each galaxy catalogue is obtained from a snapshot of a state-of-the-art hydrodynamic simulation that follows the evolution of $256^3$ dark matter particles and $256^3$ fluid elements from $z=127$ down to $z=0$. All simulations share the value of these cosmological parameters, $\Omega_{\rm b}=0.049$, $\Omega_{\rm K}=0$, $h=0.6711$, $n_s=0.9624$, $w=-1$, $\sum_i m_{\nu_i}=0$ eV, and they can be classified into two different suites:
\begin{itemize}
    \item \textbf{IllustrisTNG}. These simulations have been run with the \textsc{Arepo}\footnote{\url{https://arepo-code.org/}} code \citep{Arepo, Weinberger:2019tbd} and employ the IllustrisTNG subgrid physics model \citep{2017MNRAS.465.3291W, 2018MNRAS.473.4077P, 2019ComAC...6....2N}. 
    \item \textbf{SIMBA}. These simulations have been run with the \textsc{Gizmo}\footnote{\url{http://www.tapir.caltech.edu/~phopkins/Site/GIZMO.html}} code \citep{Hopkins2015_Gizmo} and employ the SIMBA subgrid physics model \citep{Dave:2019yyq}. 
\end{itemize}
We employ the so-called LH set of both suites, that contains 1,000 simulations with different values of the cosmological and astrophysical parameters, and initial random seed. The cosmological parameters varied in the simulations are the matter density parameter, $\Omega_{\rm m}$, and the variance of the linear field on scales of 8 $h^{-1}$Mpc at $z=0$, $\sigma_8$. The astrophysical parameters are denoted by $A_{\rm SN1}$, $A_{\rm SN2}$, $A_{\rm AGN1}$, and  $A_{\rm AGN2}$, and control the efficiency of supernovae ($A_{\rm SN1}$ and $A_{\rm SN2}$) and active galactic nuclei ($A_{\rm AGN1}$ and $A_{\rm AGN2}$) feedback. The values of these six parameters are arranged in a latin-hypercube with boundaries set by
\begin{eqnarray}
\Omega_{\rm m} &\in&[0.1 ; 0.5]\\
\sigma_8 &\in&[0.6 ; 1.0]\\
A_{\rm SN1}, A_{\rm AGN1} &\in& [0.25 ; 4.0]\\
A_{\rm SN2}, A_{\rm AGN2} &\in& [0.5 ; 2.0]~.
\end{eqnarray}
The simulations in the two suites differ not only on the value of the cosmological and astrophysical parameters, but also in the way the hydrodynamic equations are solved and in the subgrid model recipes. We emphasize that having two independent models to simulate galaxies, cosmic gas, and in general the large-scale structure of the Universe is key in order to asset the \textit{robustness} of the developed machine learning model. 

\begin{figure*}[th!]
\begin{center}
\includegraphics[width=0.325\linewidth]{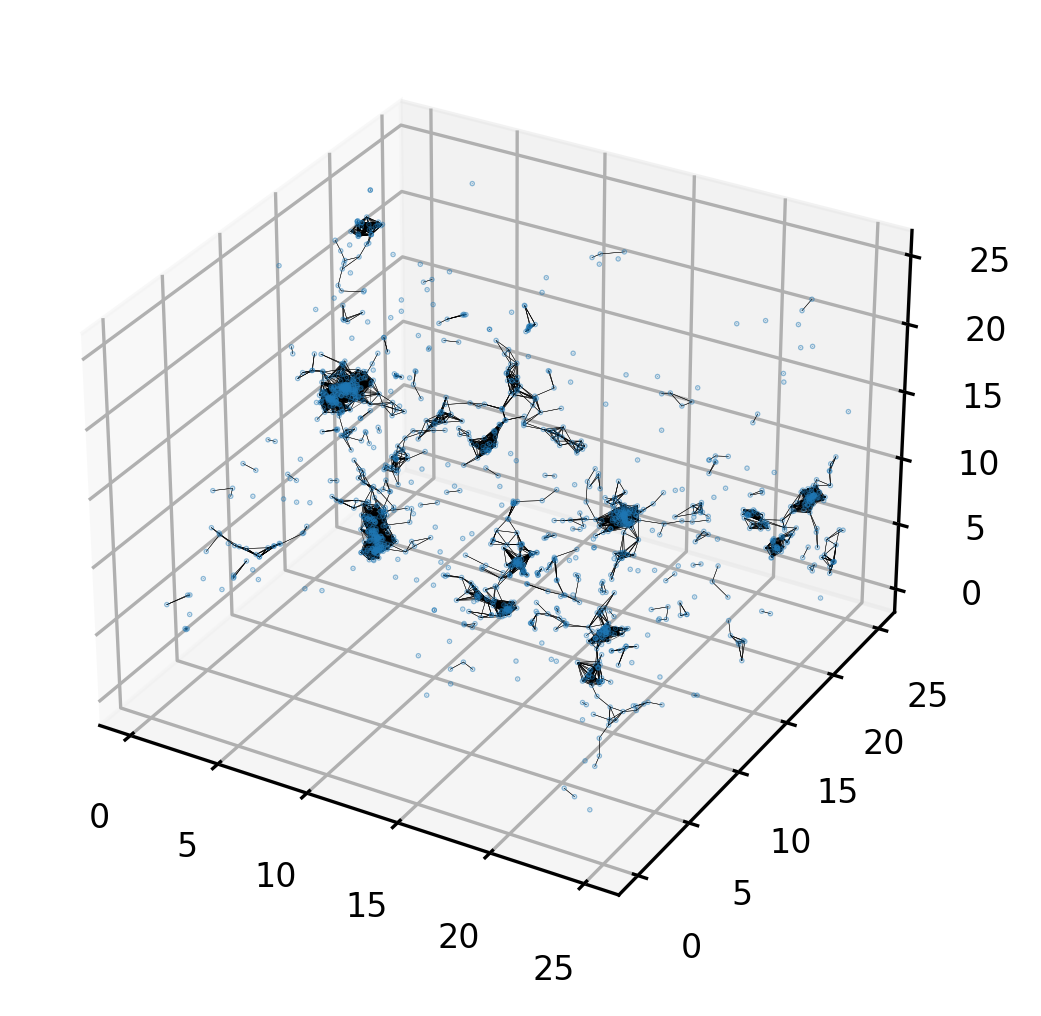}
\includegraphics[width=0.325\linewidth]{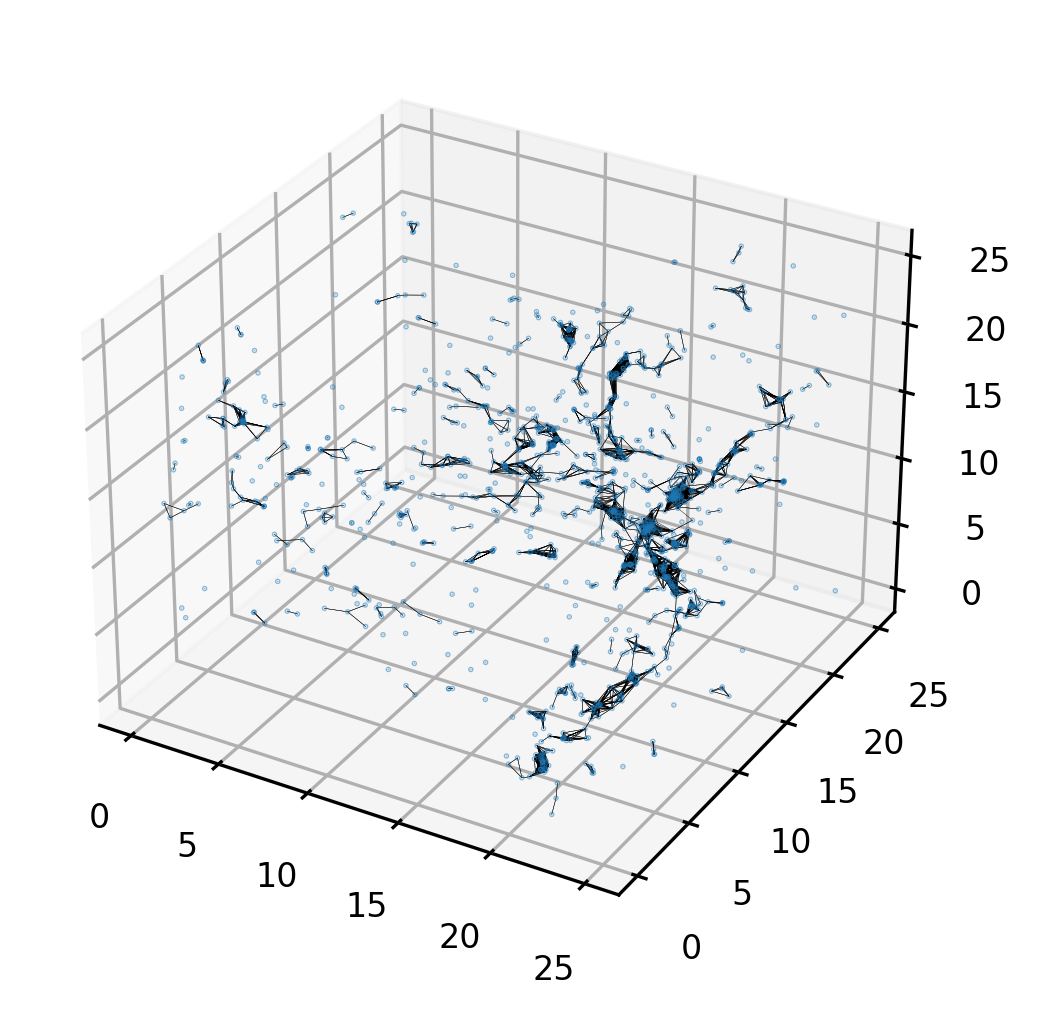}
\includegraphics[width=0.325\linewidth]{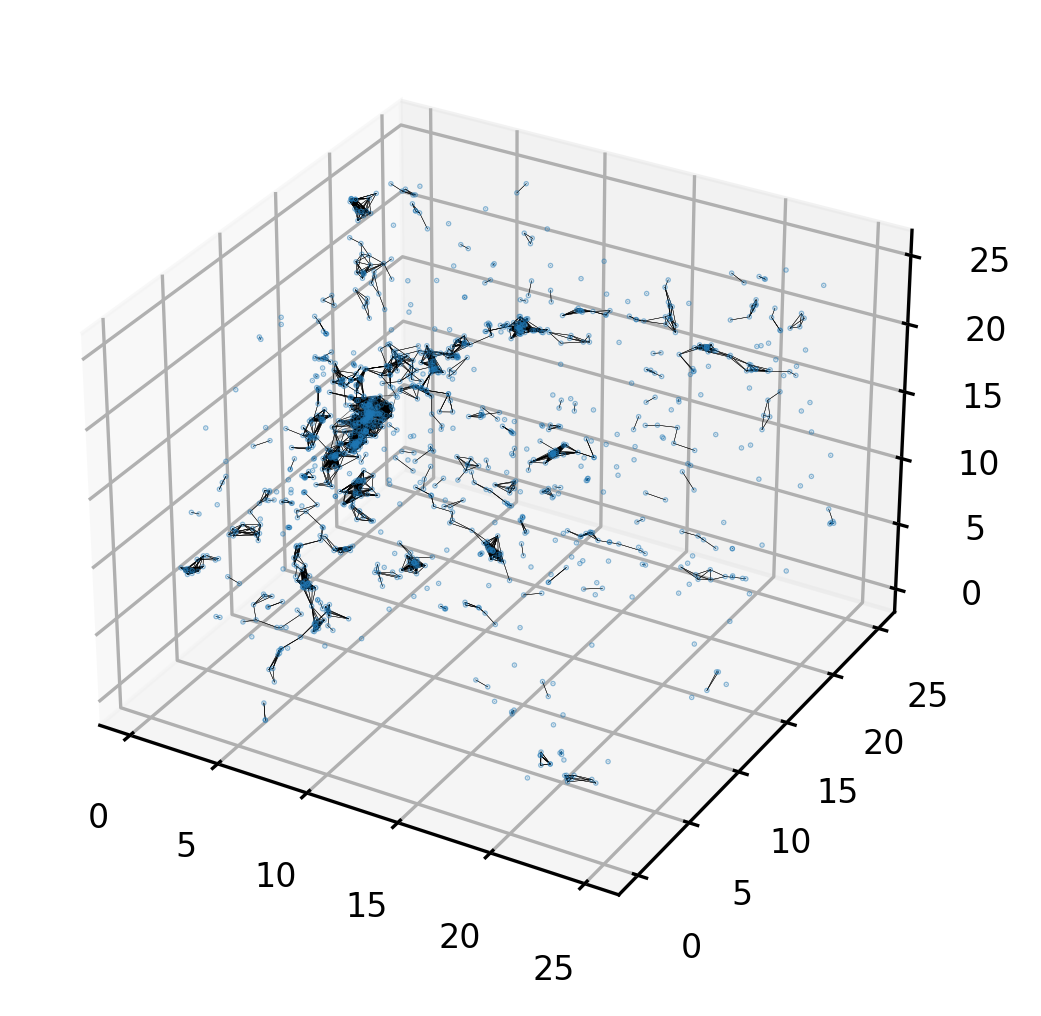}
\caption{The input to our model is a graph, that we build from a given galaxy catalogue. In the graph, the nodes represent galaxies and two nodes are linked by an edge if their distance is smaller than the linking radius $r$. This figure show three graphs from three different CAMELS-IllustrisTNG simulations where $r=1.25~h^{-1}{\rm Mpc}$. Each box represents the galaxy catalogue of a simulation at $z=0$. Box units are in $h^{-1}{\rm Mpc}$.}
\label{fig:graphs}
\end{center}
\end{figure*}

In total we have 2,000 galaxy catalogues at a given redshift: 1,000 from the IllustrisTNG simulations and 1,000 from the SIMBA simulations. At $z=0$, the galaxy catalogues from the IllustrisTNG simulations contain an average of 700 galaxies, while this number raises to 1200 for the SIMBA simulations. For each snapshot of each simulation we have identified halos and subhalos by running \textsc{SUBFIND} \citep{Subfind}. In this paper we work with galaxies, that we define as subhalos that contain at least 20 star particles Thus, our galaxy catalogues contain galaxies with stellar masses above $\sim 10^8~h^{-1}M_\odot$. We have checked that using a slightly different threshold does not change our conclusions. We use galaxy catalogues in real-space, and leave for future work the analysis for galaxy catalogues in redshift-space. While for regression we will work with galaxy catalogues at different redshifts, for inference we only consider galaxy catalogues at $z=0$. For each simulation we have thus a galaxy catalogue that contains the following information for each galaxy:
\begin{itemize}
    \item 3D position, $\textbf{p}$. The Cartesian coordinates of the galaxy in real-space.
    \item Stellar mass, $M_*$. Defined as the total mass in stars in the galaxy.
    \item Stellar radius, $R_*$. Defined as the radius that contains half of the galaxy's stellar mass. 
    \item Stellar metallicity, $Z_*$. Defined as the mass-weighted metallicity of the stars in the galaxy. 
    \item Maximum circular velocity, $V_{\rm max}$. Defined as the maximum value of $\sqrt{G M(<r)/r}$. This quantity is typically dominated by the dark matter content of a given subhalo.
\end{itemize}
We emphasize that all considered galaxy properties, with the exception of $V_{\rm max}$, are observationally accessible. The idea behind the usage of these features is that galaxy properties, individually or collectively, may 1) carry out cosmological information and/or 2) contain astrophysical information that can be used to break cosmological-astrophysical degeneracies.

To illustrate the potential information that galaxy properties may encode, we show in Fig. \ref{fig:jointplot} correlations among different galaxy features, color coded by $\Omega_m$, for galaxies in the $z=0$ catalogues. As can be seen, and already noted in \cite{2022arXiv220102202V}, there are some noticeable correlations between galaxy properties and $\Omega_{\rm m}$. Thus, our model may learn to extract cosmological information based on galaxy properties, on top of galaxy clustering. It is important to note that the trends between the different galaxy features and $\Omega_m$ can significantly vary from the IllustrisTNG suite (left) to the SIMBA suite (right). This fact illustrates the differences in the subgrid models employed in the IllustrisTNG and SIMBA simulations. We will investigate in Sec. \ref{sec:crosstest} the robustness of our results to changes in the subgrid model.

When training GNNs with galaxy catalogues, we will consider all models equally probable. For instance galaxy catalogues from simulations with very efficient supernova feedback are considering equally probable than catalogues from simulations with more realistic feedback efficiencies. This is equivalent to consider a flat prior on the value of both the cosmological and astrophysical parameters. Given the broad range in the value of the astrophysical parameters, and the fact that we are interested in inferring the value of the cosmological parameters, it is appropriate to say that our networks are trained to learn to marginalize over baryonic effects, as implemented in CAMELS.

%%%%%%%%%%%%%%%%%%%%%%%%%%%%%%%%%%%%%%%%%%%%%%%%%%%%%%%
%%%%%%%%%%%%%%%%%%%%%%%%%%%%%%%%%%%%%%%%%%%%%%%%%%%%%%%
\section{Model}
\label{sec:model}

In this section we first describe how we transform the galaxy catalogues into graphs, the input to our model. We then describe the architecture of our GNNs and explain how the loss function is set depending on the task to be carried out: regression or likelihood-free inference.

%\hfill \break

\subsection{From galaxy catalogues to cosmic graphs}

We represent each galaxy catalogue as a mathematical graph $\mathcal{G}$ that is built as follows. Galaxies represent the graph nodes, and two galaxies are connected by an edge if their separation is smaller than the linking radius $r$, an hyperparameter. We show some examples of graphs constructed from galaxy catalogues of IllustrisTNG simulations in Fig. \ref{fig:graphs}.

We note that the method to construct the graphs from the galaxy catalogues is not unique. For instance, the $k$ nearest neighbors of a given node can be used to define the graph's edges. However, we have checked that this method led to worse results than the linking radius approach. 

We now describe in detail the different elements of the graphs and what features are used in order to respect the underlying symmetries. As we shall see below, the architecture of our model uses several graph blocks, that will take a given graph and output another one with some of its attributed updated. Therefore, we will describe the graph properties at a given graph block $l$.

\subsubsection{Node features}

The features of the node $i$ at block $l$ are represented by $\textbf{h}_i^{(l)}$. The initial node features, $\textbf{h}_i^{(0)}$, are set by the intrinsic properties of the node's host galaxy introduced in Sec. \ref{sec:data}: 1) the stellar mass $M_*$, 2) the stellar half-mass radius $R_*$, 3) the stellar metallicity $Z_*$, and 4) the maximum subhalo velocity $V_{\rm max}$. We note that a subset of these properties can also be used as node features, e.g. only the stellar mass. In some cases we employ catalogues that only contain galaxy positions. In such cases, we do not use initial node features, although hidden node features are computed, as discussed in Sec. \ref{sec:arch}.

\subsubsection{Edge features}

We denote the edge features between nodes $i$ and $j$ at the block $l$ as $\textbf{e}_{ij}^{(l)}$. The edges play a crucial role in our model since they account for the spatial information. Due to statistical homogeneity and isotropy, we would like our model to be invariant under translations and rotations, i.e., rigid transformations of the Euclidean group. Let $\textbf{p}_i$ be the position of the galaxy in the node $i$. Given a rotation matrix $\textbf{R}$ and a translation $\textbf{T}$, positions transform as $\textbf{p}_i \rightarrow \textbf{R}\textbf{p}_i+\textbf{T}$. Requiring that the output of our network is invariant under these transformations impose restrictions over the spatial information to be employed.

Translational symmetry can be imposed demanding that the edge features should be given in terms of relative positions rather than the absolute ones. Using $\textbf{d}_{ij}=\textbf{p}_i-\textbf{p}_j$, instead of $\textbf{p}_i$ or $\textbf{p}_j$, will already fulfill this requirement. 

Symmetry under rotations can be built by using as features the distance and the scalar products of the directions of two vectors defining the edge \citep{2021arXiv210606610V}. Lets define the unit vectors $\textbf{s}_{ij} = \textbf{d}_{ij}/|\textbf{d}_{ij}|$ and $\textbf{n}_i = (\textbf{p}_i-\bar{\textbf{p}})/|\textbf{p}_i-\bar{\textbf{p}}|$, where $\bar{\textbf{p}}$ is the centroid of the distribution\footnote{The center of mass or the position of any galaxy could also be employed, as long as $\bar{\textbf{p}}$ transforms as $\textbf{p}_i$.}. Then, we can employ as edge features the scalar products $\alpha_{ij} = \textbf{n}_i \cdot \textbf{n}_{j}$ and $\beta_{ij} = \textbf{n}_i \cdot \textbf{s}_{ij}$, together with the relative distance, $|\textbf{d}_{ij}|$. Note that these three scalar features are equivalent to the vector  $\textbf{p}_i-\textbf{p}_j$, and therefore our model does not loose information by using these features instead of the relative distance vector.\footnote{Note that there is a third possible product, $\textbf{n}_j \cdot \textbf{s}_{ij}$, but this scalar should not provide additional independent information since the vectors $\textbf{p}_i$, $\textbf{p}_j$ and $\textbf{d}_{ij}$ form a triangle whose angles must sum $\pi$ radians. Thus, only two angles are independent, and we evaluate them using $\alpha_{ij}$ and $\beta_{ij}$. Other way to reason is that $\textbf{d}_{ij}$ carries three degrees of freedom, the same than the quantities $|\textbf{d}_{ij}|$, $\alpha_{ij}$ and $\beta_{ij}$.} Since it is better to input the model features that are dimensionless and normalized, we divide  $|\textbf{d}_{ij}|$ by the linking radius $r$. Thus, the initial edge features between nodes $i$ and $j$, $\textbf{e}_{ij}^{(0)}$, are given by
\begin{equation}
\textbf{e}_{ij}^{(0)} = [\,|\textbf{d}_{ij}|/r,\,\alpha_{ij},\,\beta_{ij}\,]~,
\label{Eq:edge_features}
\end{equation}
where $[...]$ denotes concatenation (in the \textit{features} axis). Note that by construction, $\textbf{e}_{ij}^{(0)}$ is translation and rotation invariant, i.e., rigid transformations of the form $\textbf{p}_i \rightarrow \textbf{R}\textbf{p}_i+\textbf{T}$ leave $\textbf{e}_{ij}^{(0)}$ invariant. Since all spatial information of the data is encoded in the edge features, and these are constructed to be translational and rotational invariant, the entire network will satisfy those symmetries by construction. 

We have explicitly checked that using the vector $\textbf{d}_{ij}$ and the distance as edge features leads to similar performance, although in this case data augmentation (e.g. random rotations) is needed to teach the model the rotational symmetry.

We consider self-loops when galactic features are included, but we disregard them when only using galaxy positions. For self-loops the edge attributes are set to $\alpha_{i,i}=1$ and $\beta_{i,i}=0$ for consistency.

\subsubsection{Global features}

Besides node and edge attributes, a graph can have global features, $\textbf{u}$, that characterize properties of the entire graph. These properties can be used in conjunction with the edge and/or node features to update the different components of the graph. In our case, we make use of the logarithm of the total number of galaxies in the simulation as a global feature. 

We have checked that in the case of catalogues that contain galaxy properties, using additional global properties, like the total stellar mass of the catalogue, does not improve the accuracy of the model. Thus, we do not consider additional global properties.

\begin{figure*}[th!]
\begin{center}
\includegraphics[width=0.8\linewidth]{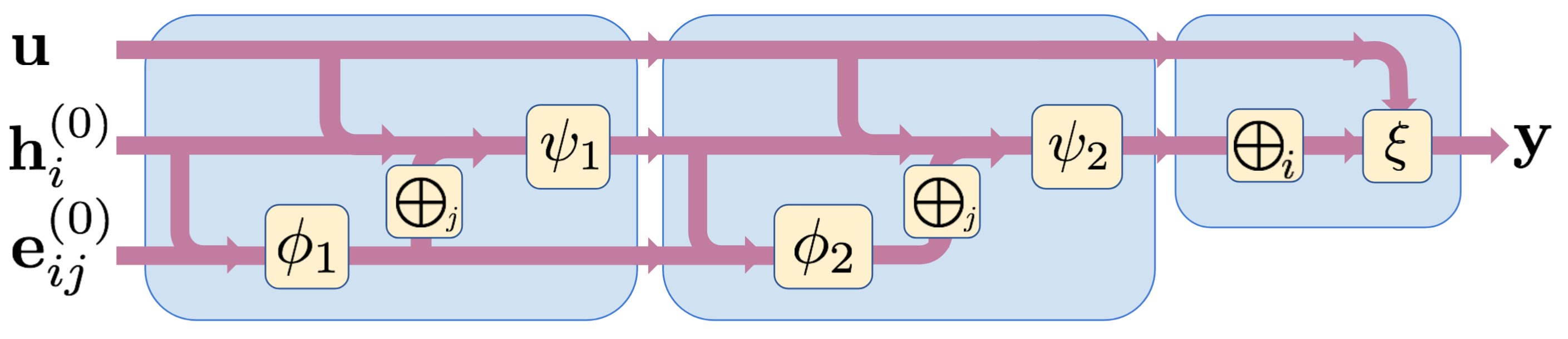}
\caption{Example of the GNN architecture with two graph blocks. Node, edge and global features are input into the network and updated through the different message passing layers. A final pooling step aggregates the information from all nodes to output the target $\textbf{y}$. }
\label{fig:model}
\end{center}
\end{figure*}

\subsection{Architecture}
\label{sec:arch}

We follow the Graph Network framework \citep{2018arXiv180601261B} to build the architecture of our model. Our GNNs are composed by different graph blocks, also called message passing layers, which update node and edge features, followed by a final aggregation layer which outputs global quantities from the full graph. In the following we shall discuss in detail how the different graph elements are updated by the graph blocks and how the final aggregation layer works. An scheme with building blocks of our GNN is shown in Fig. \ref{fig:model}. We refer the reader to \cite{2021arXiv210413478B, 2018arXiv180601261B, HamiltonBook} for general references on GNNs.

The graph block $l$ is composed by:
\begin{itemize}
\item An edge model that takes as input the edge and node features and outputs new edge features: 
\begin{equation}
    \textbf{e}_{ij}^{(l+1)}=\phi_{l+1} ([\textbf{h}_i^{(l)},\textbf{h}_j^{(l)},\textbf{e}_{ij}^{(l)}]), 
    \label{eq:edgelayer}
\end{equation}
\item A node model that takes as input the node and global features together with the updated edge features and returns new node features:
\begin{equation}
    \textbf{h}_i^{(l+1)} = \psi_{l+1} ([\textbf{h}_i^{(l)}, \bigoplus_{j \in \mathcal{N}_i} \textbf{e}_{ij}^{(l+1)}, \textbf{u}])
    \label{eq:nodelayer}
\end{equation}
\end{itemize}
In the above expressions, $\phi_{l+1}$ and $\psi_{l+1}$ represent generic functions modeled through multilayer perceptrons (MLPs), $\mathcal{N}_i$ denotes the neighborhood of node $i$, i.e., the collection of all other nodes with an edge connected to $i$, and $\bigoplus$ is a permutation invariant aggregation operator, which aggregates the information from all the neighbors of node $i$. Examples of such operators are the sum, the mean, and the maximum. In our case, we do not know a-priori which operator will perform better. Thus, we use the three of them and concatenate the results into an array:
\begin{equation}
    \bigoplus_{j \in \mathcal{N}_i} \textbf{e}_{ij}^{(l+1)}=\left[\max_{j \in \mathcal{N}_i} \textbf{e}_{ij}^{(l+1)}, \sum_{j \in \mathcal{N}_i} \textbf{e}_{ij}^{(l+1)}, \frac{ \sum_{j \in \mathcal{N}_j} \textbf{e}_{ij}^{(l+1)}} {\sum_{j \in \mathcal{N}_j}} \right].
    \label{eq:pooling}
\end{equation}
% {\rm mean}_{j \in \mathcal{N}_i}
We have checked that using this \textit{multi-pooling} operation yields slightly more accurate results than using a single pooling operation, consistently with previous works which show the advantage of combining multiple aggregators \citep{2020arXiv200405718C}.

In some situations, we disregard galactic features and only work with galaxy positions. In such cases, no initial node features are provided, and the initial graph blocks are modified to perform
\begin{equation}
    \textbf{e}_{ij}^{(1)}=\phi_1 (\textbf{e}_{ij}^{(0)} ) 
    \label{eq:edgepos}
\end{equation}
and
\begin{equation}
    \textbf{h}_i^{(1)} = \psi_1 ([\bigoplus_j \textbf{e}_{ij}^{(1)},\textbf{u}]),
    \label{eq:nodepos}
\end{equation}
while the rest of the graph blocks present the same form of Eq. \ref{eq:edgelayer} and Eq. \ref{eq:nodelayer}.

It is well known that residual layers, such as adding the input of the layer to its output, can ease the training process, leading to smoother loss functions \citep[see e.g.][]{2017arXiv171209913L}. We make use of residuals layers in all node and edge models, excluding the first graph block (due to the different dimensionality in their feature spaces). In such cases, we modify the output of the block $l+1$ as $\textbf{e}_{ij}^{(l+1)} \leftarrow \textbf{e}_{ij}^{(l+1)} + \textbf{e}_{ij}^{(l)}$ and $\textbf{h}_{i}^{(l+1)} \leftarrow \textbf{h}_{i}^{(l+1)} + \textbf{h}_{i}^{(l)}$.

The architecture of our model consists in $L$ graph blocks, where $L$ is a hyperparameter to be optimized. After the $L$ graph blocks, an additional final layer is used, which aggregates the information for each node on the graph to predict an array of outputs $\textbf{y}$,
\begin{equation}
    \textbf{y} = \xi ( [\bigoplus_{i \in \mathcal{G}} \textbf{h}_i^{(L)},\textbf{u}] ),
    \label{eq:target}
\end{equation}
where $\xi$ is another MLP, and $\bigoplus_{i \in \mathcal{G}}$ is the multi-pooling operation defined in Eq. \ref{eq:pooling}. 

Note that by construction, the node and edge models are permutation equivariant, which means that if a permutation is applied to the ordering of nodes and edges, their hidden representations are transformed in the same way, as desired in GNNs \citep{2021arXiv210413478B}. Furthermore, the output of the network $\textbf{y}$ is by construction permutation invariant. Moreover, as discussed above, our model is also translation and rotational invariant, due to the way the edge features are taken.

We emphasize that our model does not impose a cutoff on a given scale, contrary to grid-based methods such as CNNs, which are limited by the resolution of the grid/mesh. Treating the galaxy catalogue as a point cloud avoids such restriction, being able to extract information on all relevant scales. These are determined by the minimum distance between galaxies in the catalogue, which is about $\sim 5~h^{-1}$kpc at $z=0$ for both IllustrisTNG and SIMBA simulations.

The implementation of the GNNs outlined in this section, \textsc{CosmoGraphNet} \citep{pablo_villanueva_domingo_2022_6485804}, makes use of PyTorch Geometric \citep{Fey_Fast_Graph_Representation_2019} and is publicly available on  \href{https://github.com/PabloVD/CosmoGraphNet}{GitHub \faGithub}.\footnote{\url{https://github.com/PabloVD/CosmoGraphNet}}

\subsection{Loss function}
\label{sec:loss}

The loss function is chosen depending on the task to be carried out by the model. In this work we consider two main tasks:
\begin{itemize}
    \item \textbf{Regression}. We will use GNNs to predict the power spectrum of a set of galaxies. In this case, the target $\textbf{y}$ is a vector with the values of the power spectrum $P(k)$ at each wavenumber $k$, for 79 intervals in $k$ from 0.3 to 20 $h$Mpc$^{-1}$. For this task we employ the traditional mean squared loss function.  
    \item \textbf{Inference}. We will also use GNNs to perform likelihood-free inference on the value of the cosmological parameters. In this case, the network is trained to predict the posterior mean $\mu_a$ and standard deviation $\sigma_a$ of the considered parameter $a$ (either $\Omega_{\rm m}$ or $\sigma_8$):
\begin{eqnarray}
\mu_a(\mathcal{G}) &=& \int_{\theta_a} p(\theta_a | \mathcal{G}) \theta_a d\theta_a~,\\
&& \nonumber \\
\sigma_a^2(\mathcal{G}) &=& \int_{\theta_a} p(\theta_a | \mathcal{G}) (\theta_a - \mu_a)^2 d\theta_a~,
\label{Eq:mean_posterior}
\end{eqnarray}
where $\theta_a$ is the value of the considered cosmological parameter, $\mathcal{G}$ represents the graph, and $p(\theta_a|\mathcal{G})$ is the marginal posterior over the parameter $a$
\begin{equation}
p(\theta_a|\mathcal{G}) = \int_{\theta} p(\theta_1,\theta_2,...\theta_n | \mathcal{G})d\theta_1...d\theta_{a-1}d\theta_{a+1}...d\theta_n~.
\label{Eq:variance_posterior}
\end{equation}
The output of the GNN is hence $\textbf{y}=[\mu_a, \sigma_a]$. This task can be carried out by using a particular form for the loss function following \cite{moment_networks}. The specific form of the loss function we use to perform this task is
\begin{eqnarray}
\mathcal{L}&=&\log\left(\sum_{i\in{\rm batch}}\left(\theta_{a,i} - \mu_{a,i}\right)^2\right)\nonumber \\
+&&\log\left(\sum_{i\in{\rm batch}}\left(\left(\theta_{a,i} - \mu_{a,i}\right)^2 - \sigma_{a,i}^2 \right)^2\right)~,
\label{Eq:loss}
\end{eqnarray}
where indexes $i$ indicate each galaxy catalogue. We refer the reader to \cite{moment_networks} and \cite{2021arXiv210910915V} for further details on this loss function.

\end{itemize}

\subsection{Acccuracy metrics}

We consider several metrics to assess the accuracy of our models:

\begin{itemize}

\item \textbf{Mean relative error, $\epsilon$}, defined as 
\begin{equation}
    \epsilon = \frac{1}{N} \sum_i^N \frac{|y_{{\rm truth}, i} - y_{{\rm infer}, i}|}{y_{{\rm truth}, i}},
\end{equation}
where $N$ is the number of galaxy catalogues in the test set and $y_{{\rm truth}, i}$ and $y_{{\rm pred}, i}$ are the truth and predicted targets respectively. Note that for the inference task, $y_{{\rm truth}, i}$ and $y_{{\rm pred}, i}$ correspond to the true value of the parameter ($\theta_{a,i}$) and posterior mean ($\mu_{a,i}$) for a given parameter $a$ (either $\Omega_{\rm m}$ or $\sigma_8$), while for regression, they correspond to each component of the array $\textbf{y}$.

\item \textbf{Coefficient of determination, $R^2$}, defined as
\begin{equation}
    R^2 = 1 - \frac{\sum_i^N (y_{{\rm truth}, i} - y_{{\rm pred}, i})^2}{\sum_i^N (y_{{\rm truth}, i} - \overline{y}_{{\rm truth}})^2},
\end{equation}
with $\overline{y}_{{\rm truth}}$ the mean of true values. 

\item \textbf{$\chi^2$}, defined as 
\begin{equation}
\chi^2=\frac{1}{N}\sum_{i=1}^N \frac{(\theta_{a,i} - \mu_{a,i})^2}{\sigma_{a,i}^2}~.
\label{Eq:chi2}
\end{equation} 
and only used for the inference task. A value close to 1 indicates that the standard deviations are correctly predicted\footnote{This can be seen as minimizing the second term of Eq. \ref{Eq:loss}.}, while a larger (lower) value can be seen as an underestimation (overestimation) of the uncertainties.

\end{itemize}

\subsection{Hyperparameter optimization}

We perform hyperparameter optimization following a Bayesian-based procedure with the Tree Parzen Estimator \citep[TPE,][]{NIPS2011_86e8f7ab}, making use of the Python package Optuna\footnote{\url{https://optuna.readthedocs.io}} \citep{2019arXiv190710902A} to maximize the accuracy of the model. The hyperparameters we optimize are: 1) the learning rate, 2) the number of graph blocks, 3) the linking radius, and 4) the number of hidden features in each MLP.

The tuning is performed by searching the value of the hyperparameters that minimize the validation loss of the model. We perform a minimum of 30 trials, where each trial corresponds to the result of training the model with a given value of the hyperparameters.

%%%%%%%%%%%%%%%%%%%%%%%%%%%%%%%%%%%%%%%%%%%%%%%%%%%%%%%
%%%%%%%%%%%%%%%%%%%%%%%%%%%%%%%%%%%%%%%%%%%%%%%%%%%%%%%
\section{Learning clustering}
\label{sec:ps}

\begin{figure}[th]
\begin{center}
\includegraphics[width=0.99\linewidth]{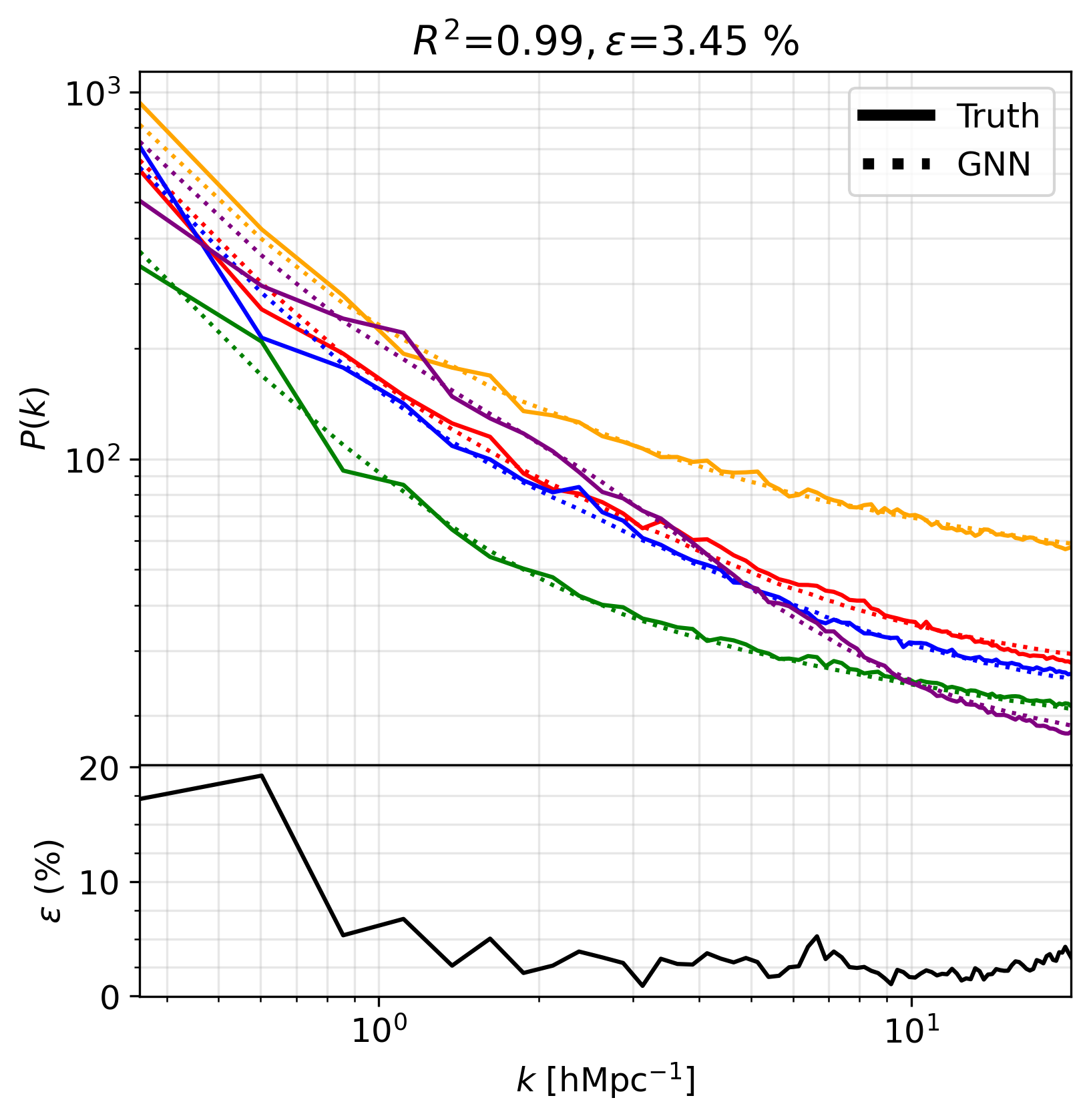}
\caption{We have trained GNNs to investigate whether they can learn to compute the power spectrum of a distribution of points, using IllustrisTNG and SIMBA galaxy catalogues at different redshifts as the training dataset. The top panel shows with solid lines power spectra from galaxy catalogues in the test set, while the predictions of the model are displayed with dotted lines. The bottom panel shows the average relative error as a function of $k$. As can be seen, the GNNs are able to learn to compute the power spectrum with a relatively high accuracy, specially at small scales.}
\label{fig:powerspec}
\end{center}
\end{figure}

We know that on sufficiently large scales, where a cosmic field resembles a Gaussian distribution, the power spectrum is the optimal estimator to extract all available information. Computing the power spectrum of a distribution of points is a non-trivial task that requires several steps\footnote{Note that there are different methods to estimate the power spectrum.}: 1) assign the points to a regular grid, 2) perform a Fourier transform, 3) compute the square of the modulus of each mode $\delta(\vec{k})^2$, and 4) compute the mean value of $\delta(\vec{k})^2$ for all modes in a given $k-$interval. This computation involves evaluating the clustering of the points on a variety of scales, from large to small. Here we want to investigate whether GNNs are able to learn to compute clustering properties of a set of points such as their power spectrum. 

We train GNNs on galaxy catalogues from both IllustrisTNG and SIMBA simulations that only contain galaxy positions. We note that in this case there is no need to separate the IllustrisTNG and SIMBA catalogues, as we are only interested in training a model that takes a distribution of points and outputs its power spectrum. For the same reason, we can also use galaxy catalogues at different redshifts simultaneously to increase our training dataset. We thus use catalogues at redshifts 0, 1, and 2. We emphasize that the graphs built from these catalogues only contain the position of the galaxies, that are embedded in the edge attributes. Hence, these graphs do not contain initial node features; we thus made use of Eqs. \ref{eq:edgepos} and \ref{eq:nodepos} for the first graph block, as explained in Sec \ref{sec:arch}. The target of our network $\textbf{y}$ is an array including the values of the power spectrum $P(k)$ at each wavenumber $k$, for 79 values of $k$ from 0.3 to 20 $h$Mpc$^{-1}$.\footnote{Note however than even smaller scales can be accessed through our graphs.}.

Top panel of Fig. \ref{fig:powerspec} shows the target (solid lines) and predicted (dotted lines) power spectra for 5 different catalogues from CAMELS simulations. We find that our model accurately predicts the galaxy power spectrum with a mean relative error of $\epsilon \simeq 3.5$\% and a $R^2$ coefficient of 0.99. The bottom panel depicts the mean relative error (averaged over the full test dataset) as a function of $k$. As can be seen, our model is more precise on small scales, where the relative error can be as small as 2.5\%, while on larger scales it increases up to 20\%. This trend can be explained taking into account that the network seems to have learned to compute a smoothed version of the power spectrum. Because of cosmic variance effects, the power spectrum is expected to be smoother on small scales and noisier on large scales, explaining why the accuracy decreases on large scales.

We have tried to improve the accuracy of our model by training GNNs with loss functions that give more weigh to the amplitude of the power spectrum on large scales. However, we did not find a significant increase in the accuracy of the model. We however believe that our results can be improved in many different ways, like using a larger dataset for training, using a more complex model, and using data with very different clustering properties on different scales. We leave all this for future work.

Nonetheless, these results show that GNNs are capable of learning to compute the clustering properties of a distribution of points. However, while this task has been carried out using galaxy catalogues of CAMELS simulations, it is not clear how well it could extrapolate when using it to compute the power spectrum from an arbitrary distribution of points. We have evaluated the extrapolation performance of this model, trained in CAMELS, by testing it on several synthetic catalogues with different clustering properties. We find that in general, the GNN fails to reproduce the correct amplitude of the power spectrum on large scales of the synthetic catalogues, while it works much better on small scales. We provide further details on this test in the Appendix \ref{sec:pstest}.

It is interesting to see that the hyperparameter optimization method selected models that only contain one graph block with a large linking radius. Surprisingly, the linking radii in these models is around $1~h^{-1}$Mpc (the value used in Fig. \ref{fig:powerspec}), that produces graphs where some nodes are isolated and therefore the model can not be using information from all galaxies when performing the regression. We believe that these models work because the GNN exploits the correlations between large and small scales to infer the former from the latter. If most of the information would be located on large scales, our model would prefer a large linking length and will avoid isolated nodes. This would be the case if our galaxy distribution would follow a Gaussian distribution, whose scales are uncorrelated. However, in this case, most of the information is located on small scales, which have a larger signal-to-noise ratio, and the graph is constructed to exploit that. While larger linking radii can slightly improve the relative error down to $\sim 2$ \%, they lead to worse performance for the extrapolation test in synthetic distributions described in Appendix \ref{sec:pstest}. Note also that larger radii imply denser graphs (with more edges), that increase the memory needed to store and manipulate them.

%%%%%%%%%%%%%%%%%%%%%%%%%%%%%%%%%%%%%%%%%%%%%%%%%%%%%%%
%%%%%%%%%%%%%%%%%%%%%%%%%%%%%%%%%%%%%%%%%%%%%%%%%%%%%%%
\section{Learning cosmology from galaxies}
\label{sec:cosmology_results}

In the previous section we showed that GNNs can learn to compute clustering properties of point distributions. Since we do not know what is the optimal estimator to extract the maximum cosmological information from galaxy catalogues, we aim to use GNNs to perform such task. Our GNNs are trained to perform field-level likelihood-free inference on the value of the cosmological parameters while marginalizing over astrophysical uncertainties, as modelled in the CAMELS simulations. 

We first trained our GNNs to infer simultaneously $\Omega_{\rm m}$ and $\sigma_8$, finding that the constraints on $\sigma_8$ were very loose. We then trained models to infer $\Omega_{\rm m}$ and $\sigma_8$ separately, and found that the bounds on $\Omega_{\rm m}$ improved significantly. Thus, from now on we only report results from models that are trained to infer the value of $\Omega_{\rm m}$ alone. The results for $\sigma_8$ are shown in the Appendix \ref{sec:sigma_8}. 

We start showing the results we obtain from galaxy catalogues that only contain galaxy positions. We then investigate how much information is gained by considering different galaxy properties together with galaxy positions. Next, the present the results of the tests performed to check the robustness of the models.

\begin{figure*}[th!]
\begin{center}
\includegraphics[width=0.45\linewidth]{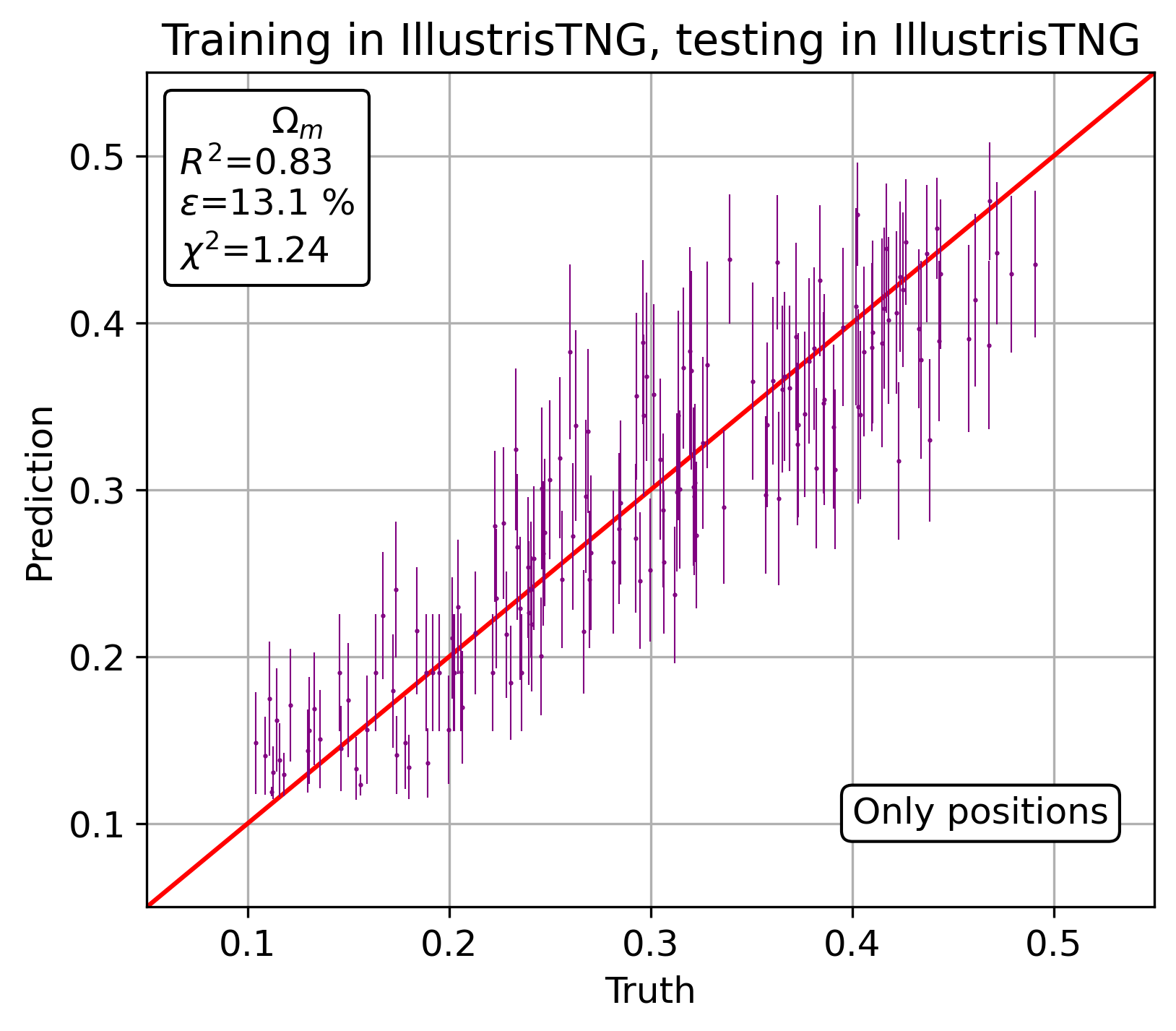}
\includegraphics[width=0.45\linewidth]{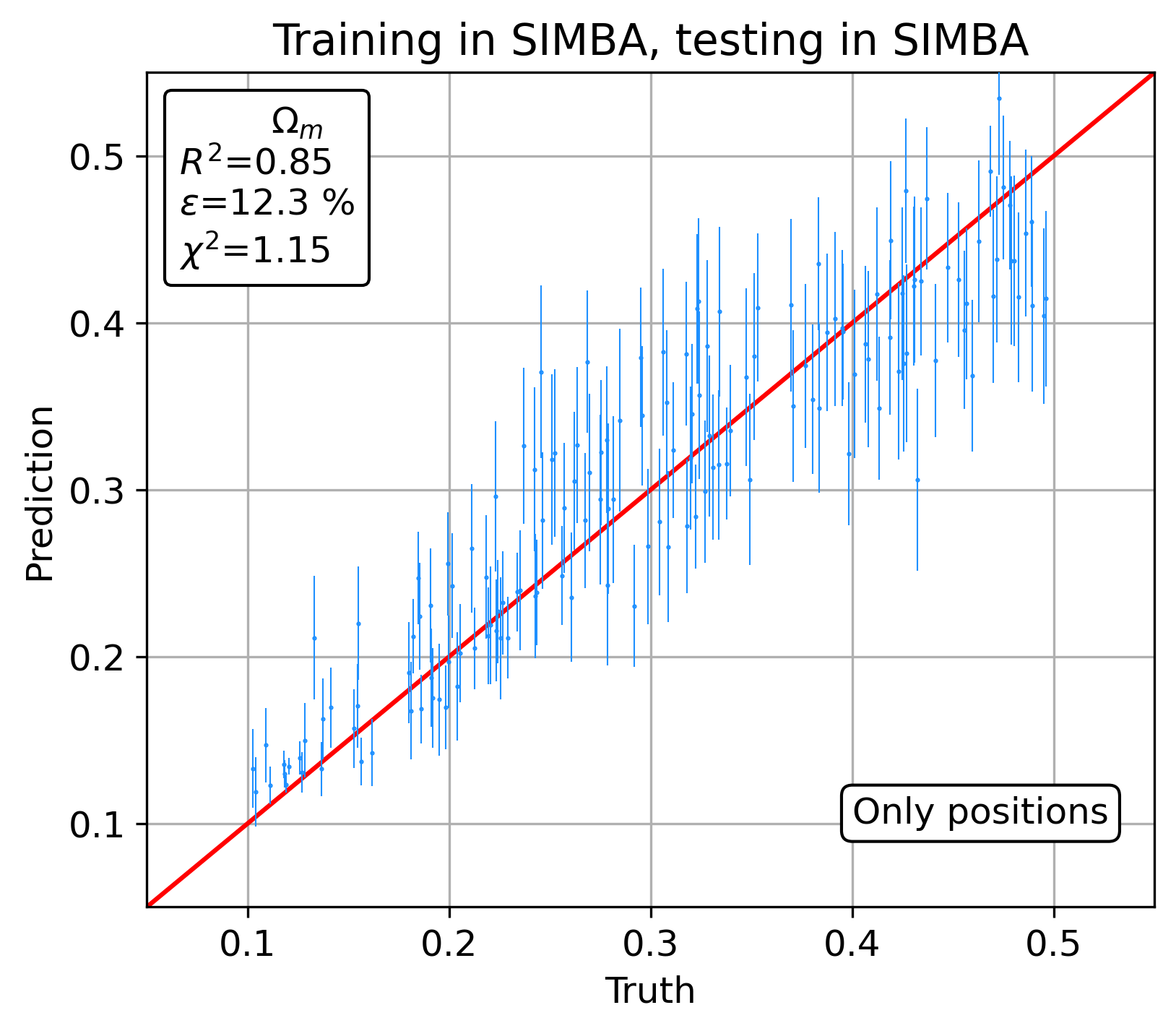}
\includegraphics[width=0.45\linewidth]{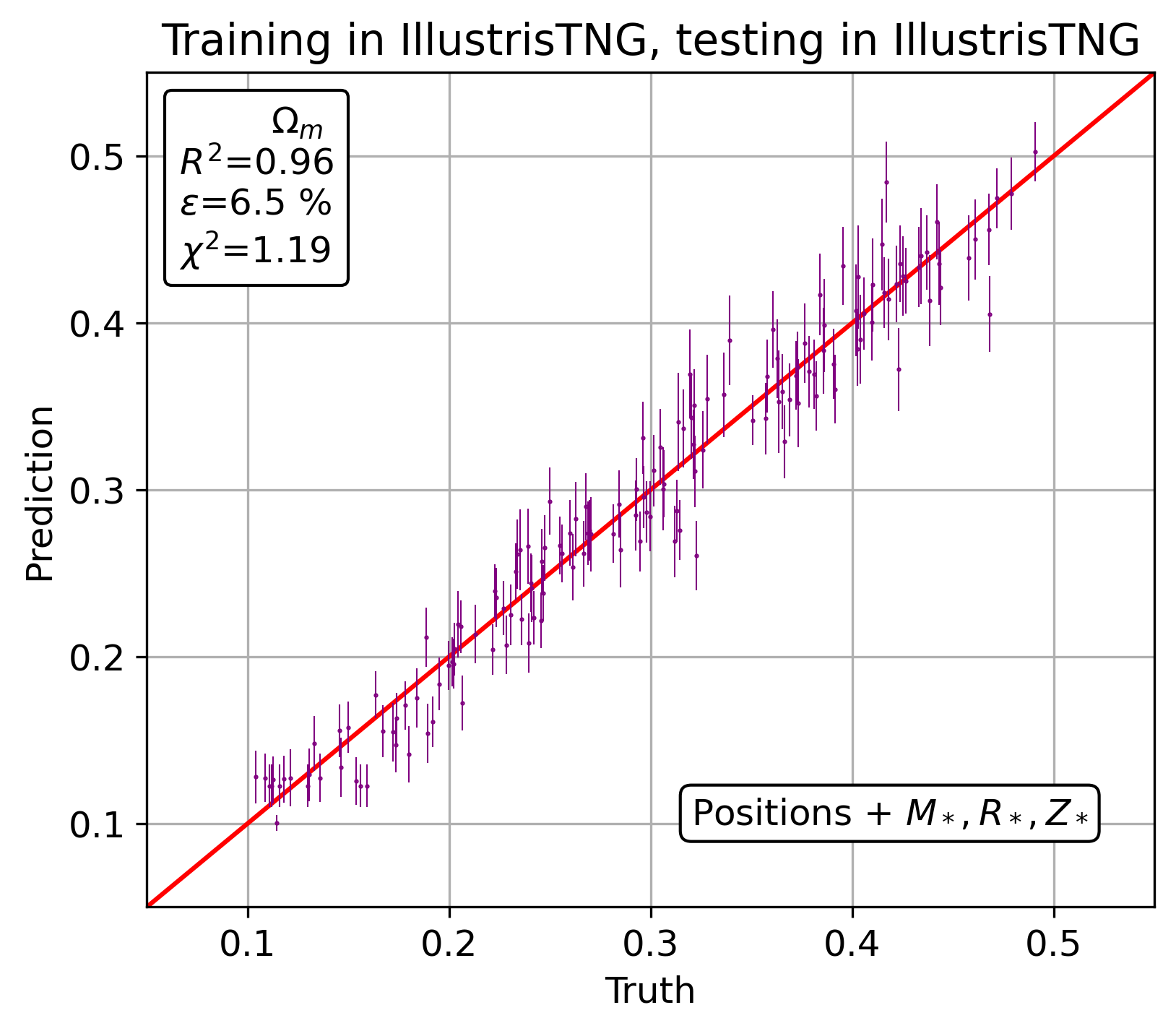}
\includegraphics[width=0.45\linewidth]{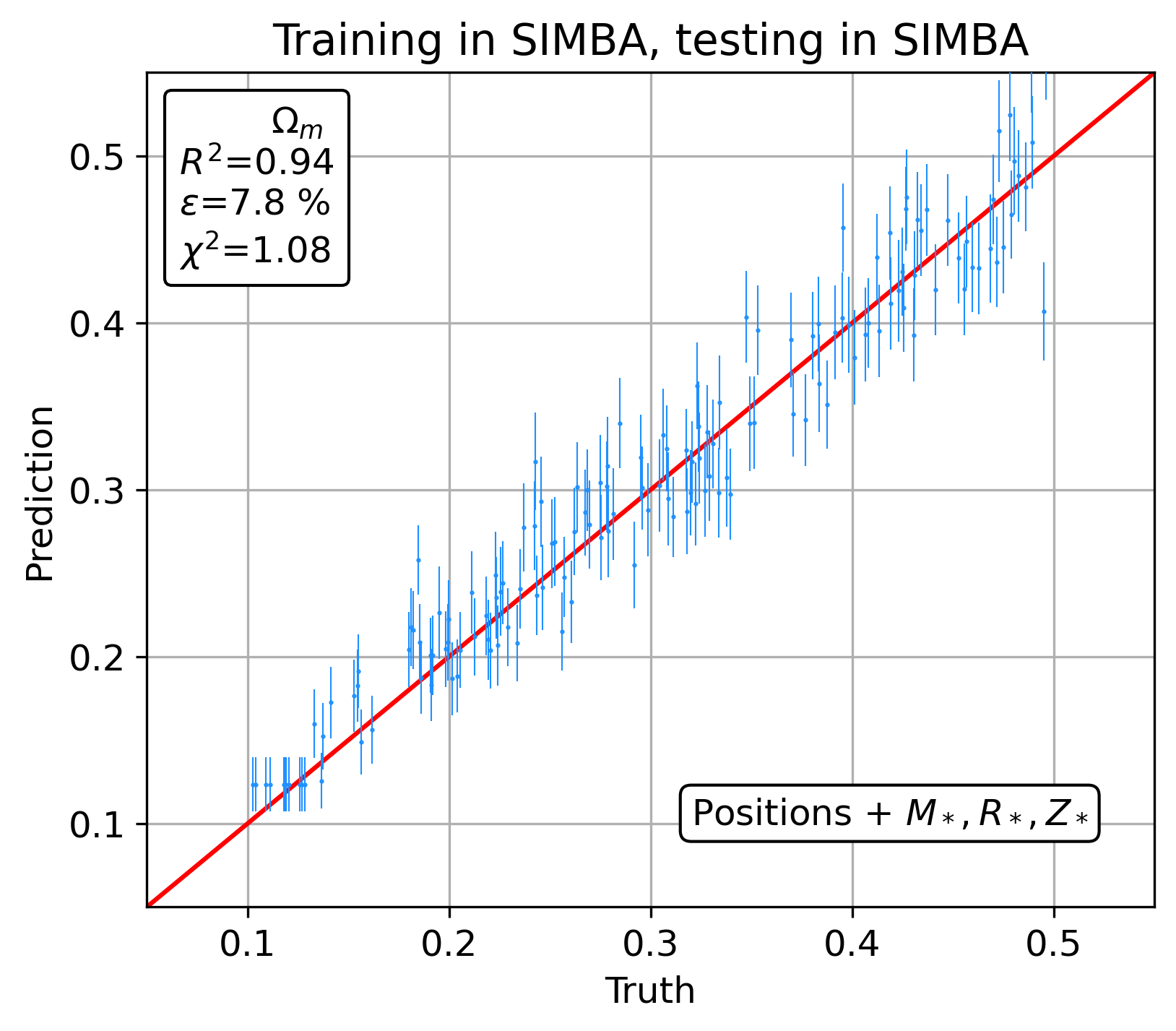}
\includegraphics[width=0.45\linewidth]{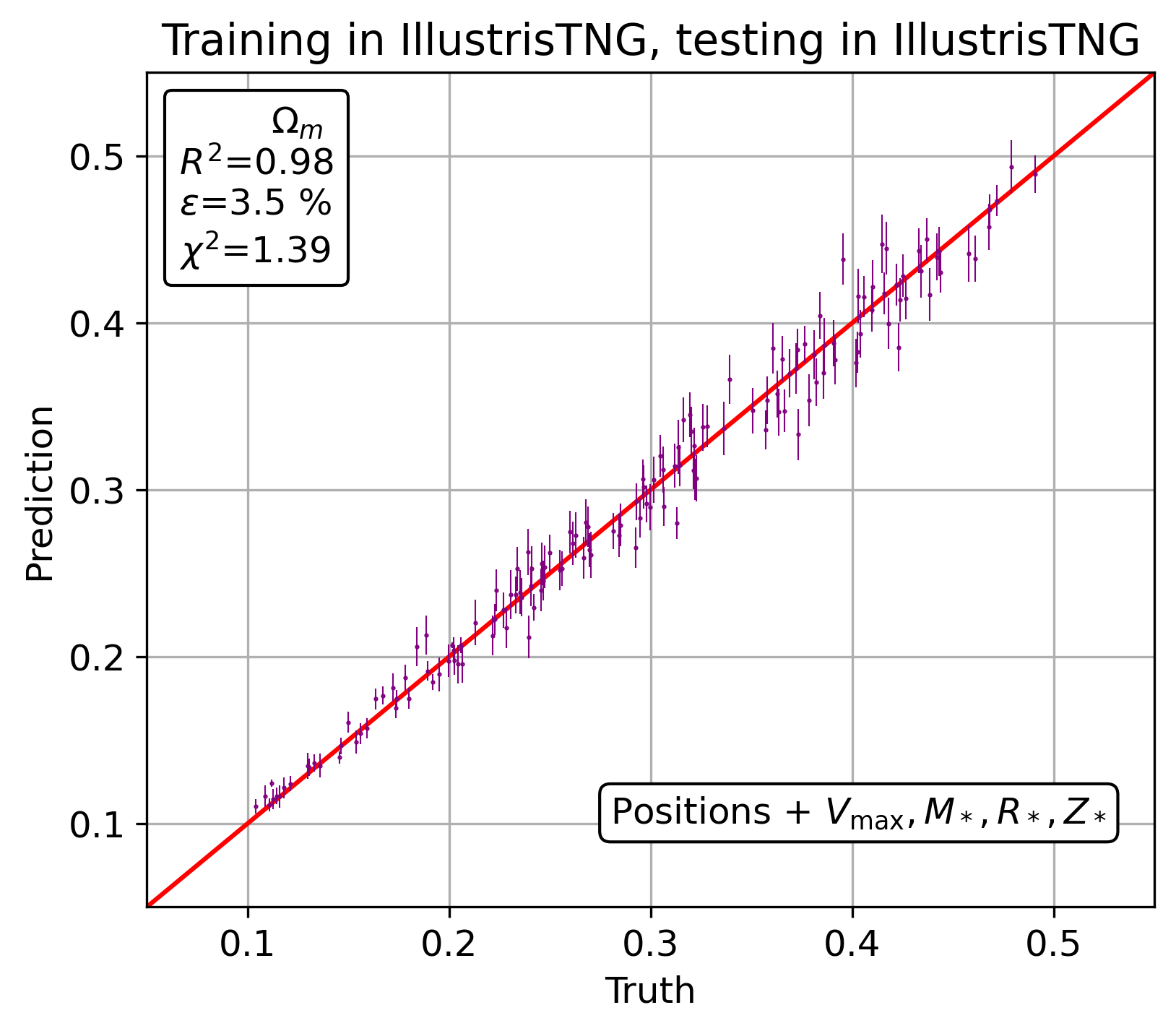}
\includegraphics[width=0.45\linewidth]{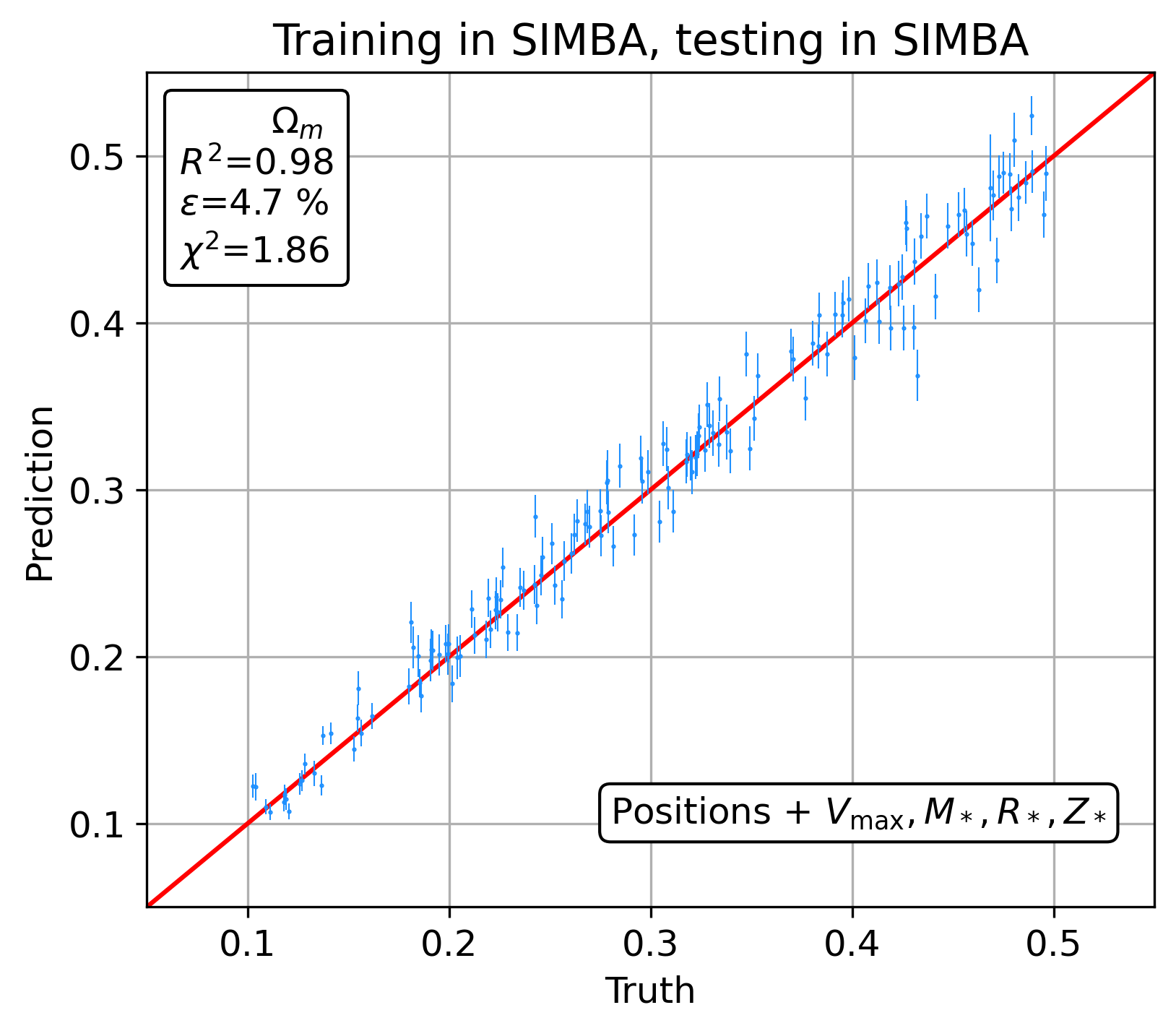}
\caption{We have trained GNNs to perform likelihood-free inference on the value of $\Omega_{\rm m}$ from galaxy catalogues that contain: 1) only galaxy positions (top row), 2) galaxy positions, stellar masses, radii, and metallicities (middle row), and 3) galaxy positions, stellar masses, radii, metallicities, and $V_{\rm max}$ (bottom row). The left column show the results of the models that have been trained and tested on catalogues from IllustrisTNG simulations, while the right column displays the equivalent for the SIMBA simulations. The numbers in the legend indicate the value of $R^2$, the mean relative error ($\epsilon$), and $\chi^2$ as defined in Eq. \ref{Eq:chi2}. As can be seen, in all cases our models are able to infer the value of $\Omega_{\rm m}$ accurately. As expected, when considering more galaxy properties, the constraints improve.}
\label{fig:Om}
\end{center}
\end{figure*}

\subsection{From galaxy positions to $\Omega_{\rm m}$}
\label{sec:posonly}

We start by investigating whether it is possible to infer the value of $\Omega_{\rm m}$ from the 3D spatial distribution of galaxies. We train and test the models using galaxy catalogues that only include galaxy positions from the IllustrisTNG and SIMBA simulations. Since this is the only property these catalogues contain, the cosmological information can only be extracted from the clustering properties of the galaxies, which can be learned in principle by a GNN, as shown in Sec. \ref{sec:ps}. The results are shown in the top row of Fig. \ref{fig:Om}. We find that the models are able to infer the value of $\Omega_{\rm m}$ with a mean relative error of $\epsilon \sim 12-13\%$ and a coefficient of determination of $R^2 \simeq 0.83-0.85$. The $\chi^2$ has values close to unity, $\simeq 1.2$, indicating that error bars are accurately predicted.

It is interesting to look at the preferred architectures found through hyperparameter optimization, that shows that the linking radius has the largest impact on the performance, preferring values between $0.1$ and $0.4~h^{-1}{\rm Mpc}$, while the specific value of the number of graph blocks and hidden channels present a milder effect. This relatively small value of the linking radius will leave some nodes without edges (see Fig. \ref{fig:graphs}). We believe that this may be happening because most of the information resides on small scales; a large value of $r$ may mix information from small with large scales in a suboptimal manner.

There are other works in the literature that have performed similar tasks. We note however that a comparison with those works is difficult due to the very different setups:
\begin{itemize}
    \item \cite{Ntampaka_19} employed convolutional neural networks to constrain $\Omega_m$ from the 3D galaxy distributions with an accuracy of $\sim 4$\%. However, their galaxy number density is lower than ours and their testing volume is more than 4000 times larger than ours. They also use galaxy catalogues obtained from HODs rather than full hydrodynamic simulations. On the other hand they incorporate effects from supersample covariance that we miss and, contrary to us, they were able to constrain $\sigma_8$.
    \item \cite{2022arXiv220402408P} performed likelihood-free inference on several summary statistics from galaxy catalogues obtained from semi-analytics models. Their constraints on $\Omega_{\rm m}$ are better than ours, but their effect volume is also much larger, with simulations boxes of 100 $h^{-1}$Mpc of length.
\end{itemize}

\subsubsection{Interpretability}

Understanding where the information to infer the cosmological parameter comes from is not obvious. We have performed several tests to gain some physical intuition on its origin.

\begin{itemize}
    \item The number of galaxies in a catalog could be expected to be correlated to the value of $\Omega_m$. However, while we have checked that there is a trend showing that simulations with higher values of $\Omega_m$ contain more galaxies, the scatter is too large to be the reason that would explain our results. Therefore, the GNN must be employing some clustering information.
    
    \item We can compare the constraints our model is placing on $\Omega_{\rm m}$ versus the ones that can be set using standard summary statistics like the power spectrum. We have quantified how much information on $\Omega_{\rm m}$ can be extracted from the galaxy power spectrum by training a multilayer perceptron that takes as input the galaxy power spectrum computed with \textsc{Pylians}\footnote{\url{https://github.com/franciscovillaescusa/Pylians3}} \citep{2018ascl.soft11008V} and returns the posterior mean and standard deviation for $\Omega_{\rm m}$. We find that the best model (after optimizing the hyperparameters) provides a mean relative error of $\epsilon \sim 30 \%$ and a correlation coefficient between the ground truth and the output of $R^2\simeq 0.5$. We note that in this case the network is almost predicting the mean of the entire distribution with large errorbars, so this relative error of $30\%$ is already heavily affected by the distribution priors. Thus, the gain from the power spectrum to the field-level inference is likely much larger than a factor of 3. This test illustrates how field-level inference yields much tighter constraints on the value of the cosmological parameters than the ones obtained from summary statistics that are suboptimal, like the power spectrum for non-Gaussian density fields.

    \item Summary statistics related to distribution and number of neighbors are known to carry out a large amount of cosmological and clustering information \citep{Banerjee_2020, Banerjee_2021}.  As an additional test, given a linking radius, we have computed the degree for each node, i.e., the number of neighbors, and generated an array of the degree distribution for each catalog. Then, a standard MLP has been trained to infer $\Omega_m$. After tunnning the hyperparameters (including the MLP depth and the linking radius), the best model can achieve a mean relative error of $\epsilon \sim 18 \%$ (with $R^2\simeq 0.74$). This result illustrates how connectivity between galaxies carries meaningful cosmological information. Indeed, it outperforms the previous MLP trained with the galaxy power spectrum. Nonetheless, a GNN can exploit clustering information in a more meaningful way to improve the degree distribution MLP, achieving $\sim 35$\% higher accuracy.
\end{itemize}

From the above tests, it seems that the GNN is able to extract non-trivial clustering information which is absent in traditional and non-standard summary statistics, obtaining a better accuracy at inferring the cosmological parameter.

\subsection{From galaxy positions and intrinsic features to $\Omega_{\rm m}$}
\label{sec:galacfeat}

In the previous subsection we have shown how the clustering of galaxies can be used to infer the value of $\Omega_{\rm m}$. We now study how much cosmological information can be extracted if we use both the clustering of galaxies and their internal properties. For this, we made use of galaxy catalogues that contain 1) galaxy positions, 2) stellar mass $M_*$, 3) stellar radius $R_*$, 4) stellar metallicity $Z_*$, and/or 5) maximum circular velocity $V_{\rm max}$. 

We have trained GNNs using galaxy catalogues from both the IllustrisTNG and SIMBA simulations. The results of testing the models with catalogues from the same simulation suites are shown in the bottom row of Fig. \ref{fig:Om}. Our networks can accurately predict $\Omega_{\rm m}$ in both simulation suites within a mean relative error of $\epsilon \simeq 3.5- 4.7\%$ and with a coefficient of determination of $R^2 \simeq 0.98$. From the $\chi^2$ values we can see that the error bars for the model trained on SIMBA can be slightly underpredicted.

In terms of hyperparameter optimization, we find that the best models prefer small number of hidden neurons (64) and linking radii of $\sim 0.4~h^{-1}{\rm Mpc}$, which is the most relevant hyperparameter, while the number of graph blocks does not have a significant impact. 

To further understand where does the information comes from, we have performed several tests:
\begin{itemize}

    \item In order to quantify how much information comes from the galaxy features versus galaxy clustering, we have trained a DeepSet \citep{2017arXiv170306114Z}, a simplified version of a GNN, which does not include neither edges nor message passing layers. It only employs node features, that are updated according to $\textbf{h}_i^{(l+1)} = \psi_{l+1} (\textbf{h}_i^{(l)})$ (instead of Eq. \ref{eq:nodelayer}). With that architecture, we find the accuracy of the model to reach mean errors of $\epsilon \sim 4-5$\%. Since this model, by construction, cannot extract information from clustering, we conclude that galaxy features are responsible for most of the information extracted from the GNNs.  
    
    \item We have trained models using only one out of the four galactic properties. We find that the most important feature appears to be the maximum circular velocity. Using $V_{\rm max}$ as the only node feature leads to accurate results, with mean relative errors around $5$ \% and $R^2 \simeq 0.97$. We note that \cite{2022arXiv220102202V} found $V_{\rm max}$ to be the most relevant feature when inferring the value of $\Omega_{\rm m}$ from individual galaxies. This could be due to the fact that $V_{\rm max}$ measures the depth of the gravitational potential, that may correlate better with $\Omega_{\rm m}$ than the other considered galaxy properties.
    
    \item We find that models trained using $M_*$, $R_*$, and $Z_*$ as galaxy properties are still accurate, with mean relative errors between $6-8$ \% and $R^2 \simeq 0.94-0.96$. This indicates that the stellar features also exhibit a strong correlation with $\Omega_m$. These results are shown in the middle panel of Fig. \ref{fig:Om}. This choice of galactic features presents an additional benefit since, contrary to $V_{\rm max}$, they are easily accessible from observational means, being more suitable to apply to real data.

\end{itemize}

To further assess the accuracy of our procedure, we can compare our results with previous works in the literature where $\Omega_m$ is predicted from cosmological simulations. To establish a fair comparison, we limit this to studies using CAMELS simulations:
\begin{itemize}
    \item In \cite{2021arXiv210910360V}, the authors were able to infer the value of $\Omega_{\rm m}$ with an accuracy between 3.4\% and 4.5\% when using 2D total mass maps. Our results have thus comparable performance, although our analysis 1) takes place in 3D instead of 2D, 2) uses galaxies instead of the full underlying matter field.
    \item In \cite{2021arXiv210909747V}, a similar approach was followed employing 13 cosmological fields. Those models reached accuracies around $2.5$ \%. We emphasize that while galaxies are directly observable, some of these 2D fields are not, or there are regions of them (like in voids and filaments) where observations will be highly affected by noise.
    \item  In \cite{2022arXiv220102202V}, $\Omega_m$ was inferred from the properties of individual galaxies. The accuracy of their results is quantified with the root-mean-squared error, $\sqrt{\langle (\Omega_{m,{\rm truth}} - \Omega_{m,{\rm pred}} )^2\rangle}$, resulting in values of $3.4 \times 10^{-2}$ and $3.7 \times 10^{-2}$ for IllustrisTNG and SIMBA respectively. Using the same metric, our model outperforms those results, with root-mean-squared errors of $1.4 \times 10^{-2}$ for both IllustrisTNG and SIMBA. We note however that our GNN model uses only four intrinsic galactic properties, instead of the 17 features assumed in \cite{2022arXiv220102202V}.
\end{itemize}

The above comparisons show that while one would expect a large gain in information by going from 2D to 3D, the usage of discrete and sparse tracers, instead of the continuous underlying field, would degrade that information. The combination of both effects yields similar results in 3D for galaxies than 2D for continuous fields.

\subsection{Robustness tests}
\label{sec:crosstest}

\begin{figure*}[th!]
\begin{center}
\includegraphics[width=0.45\linewidth]{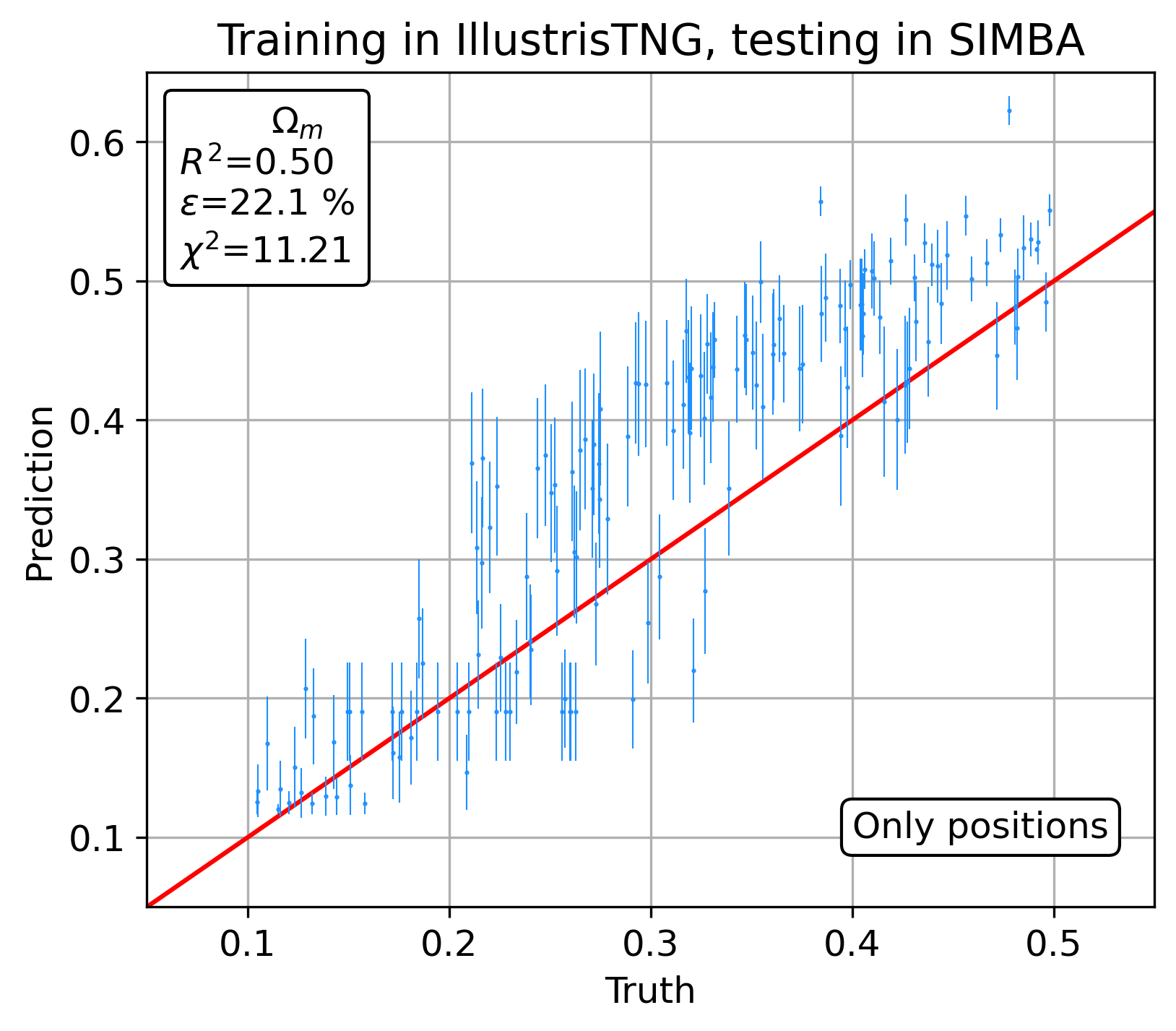}
\includegraphics[width=0.45\linewidth]{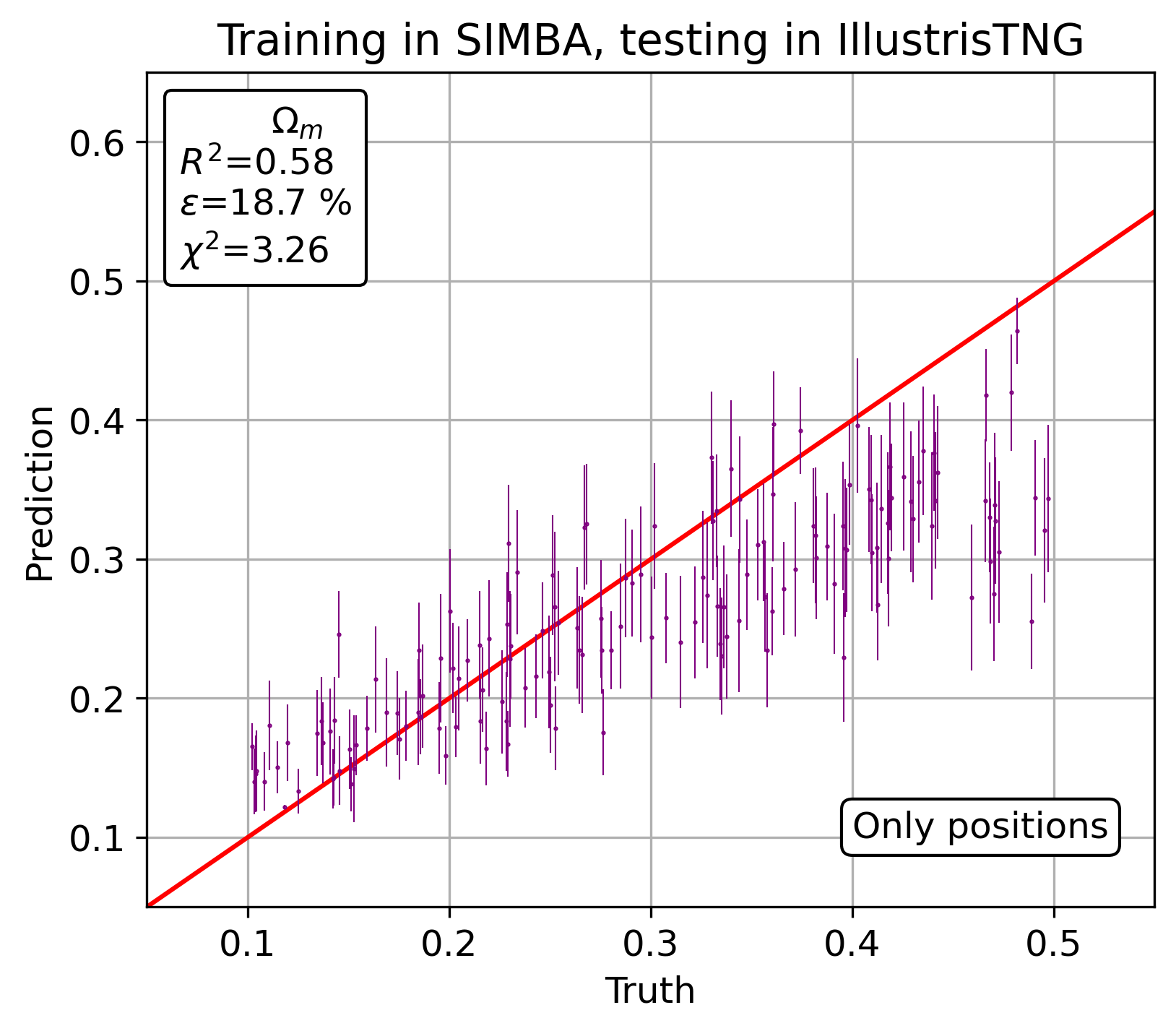}
\includegraphics[width=0.45\linewidth]{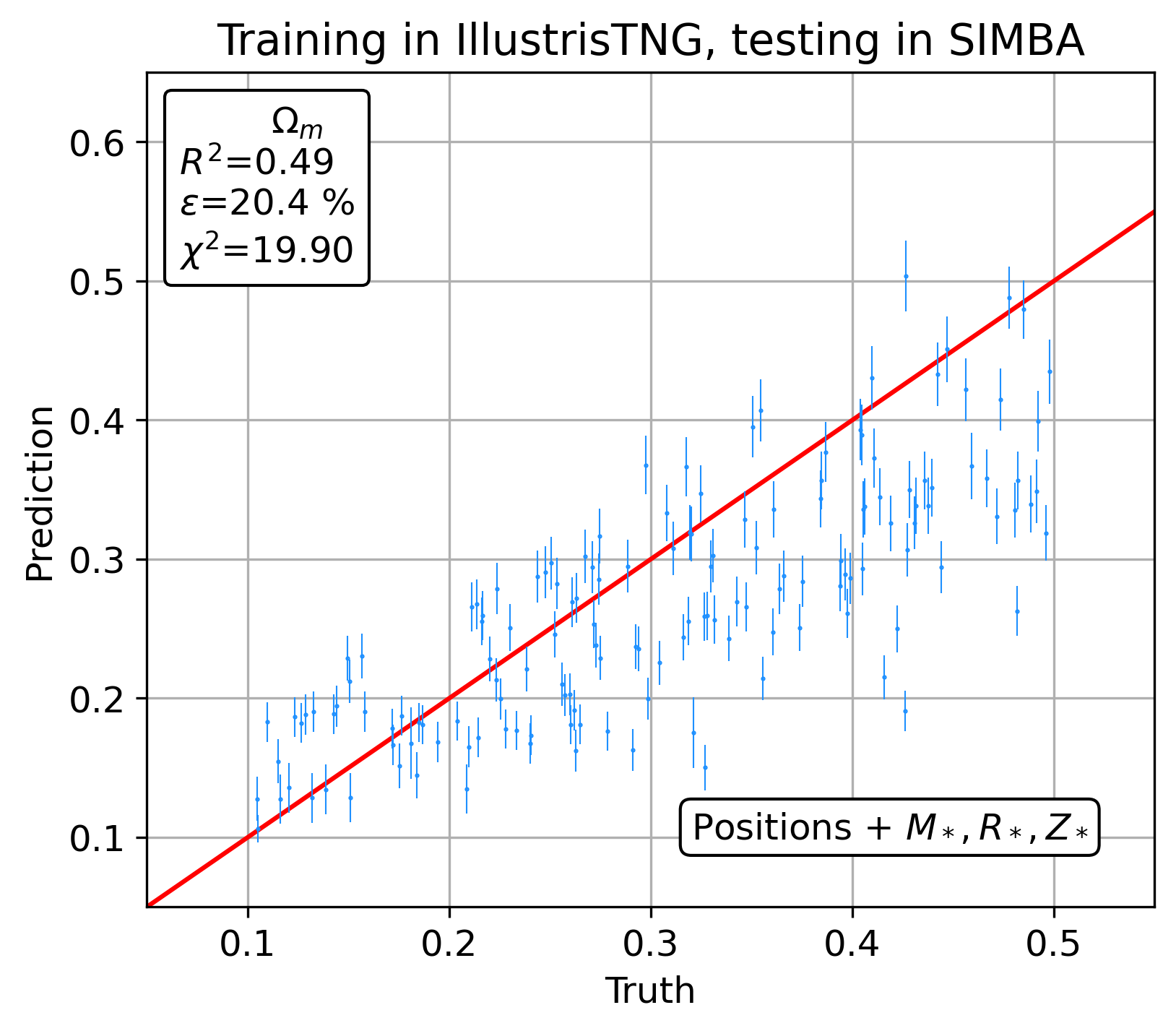}
\includegraphics[width=0.45\linewidth]{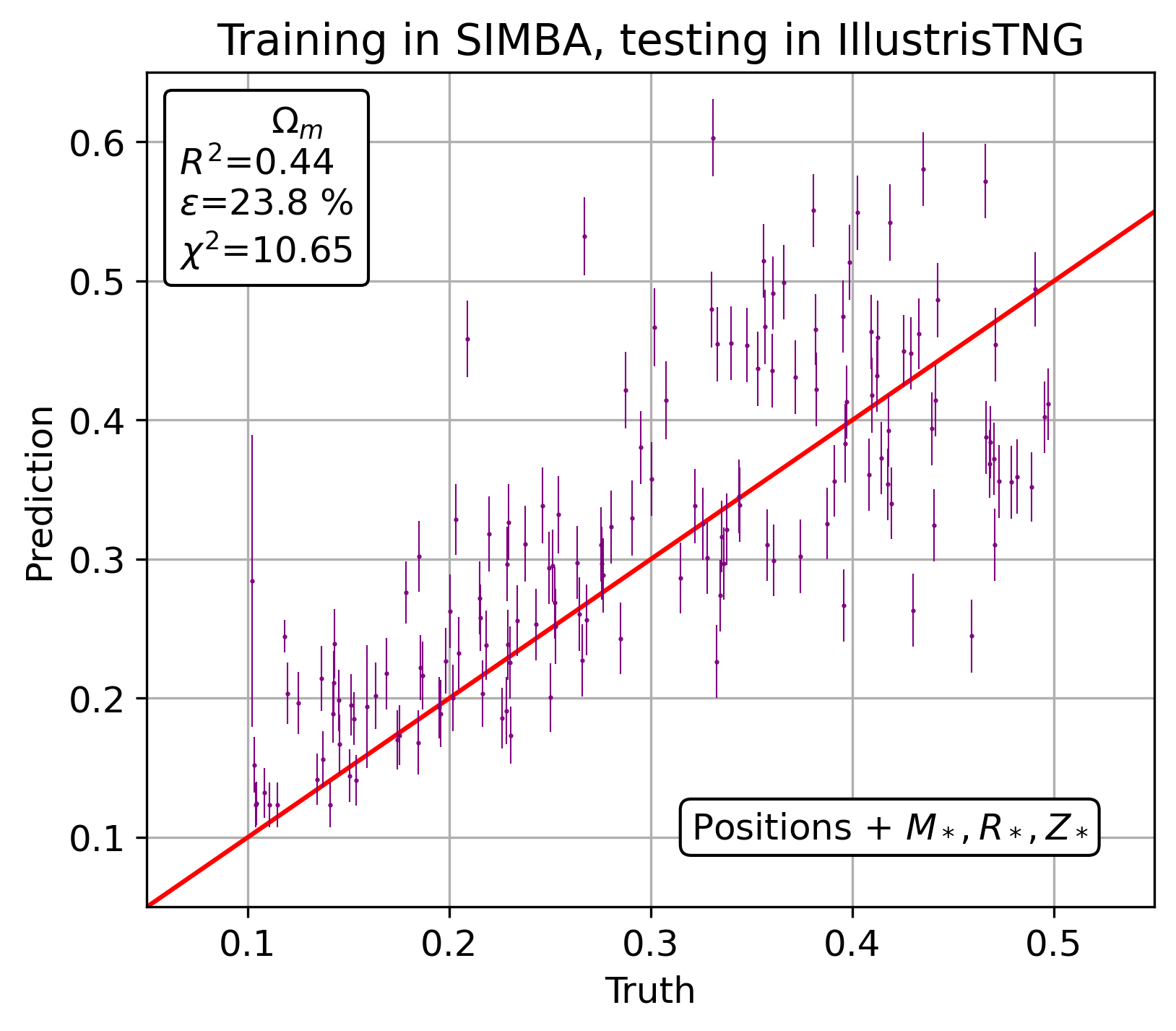}
\includegraphics[width=0.45\linewidth]{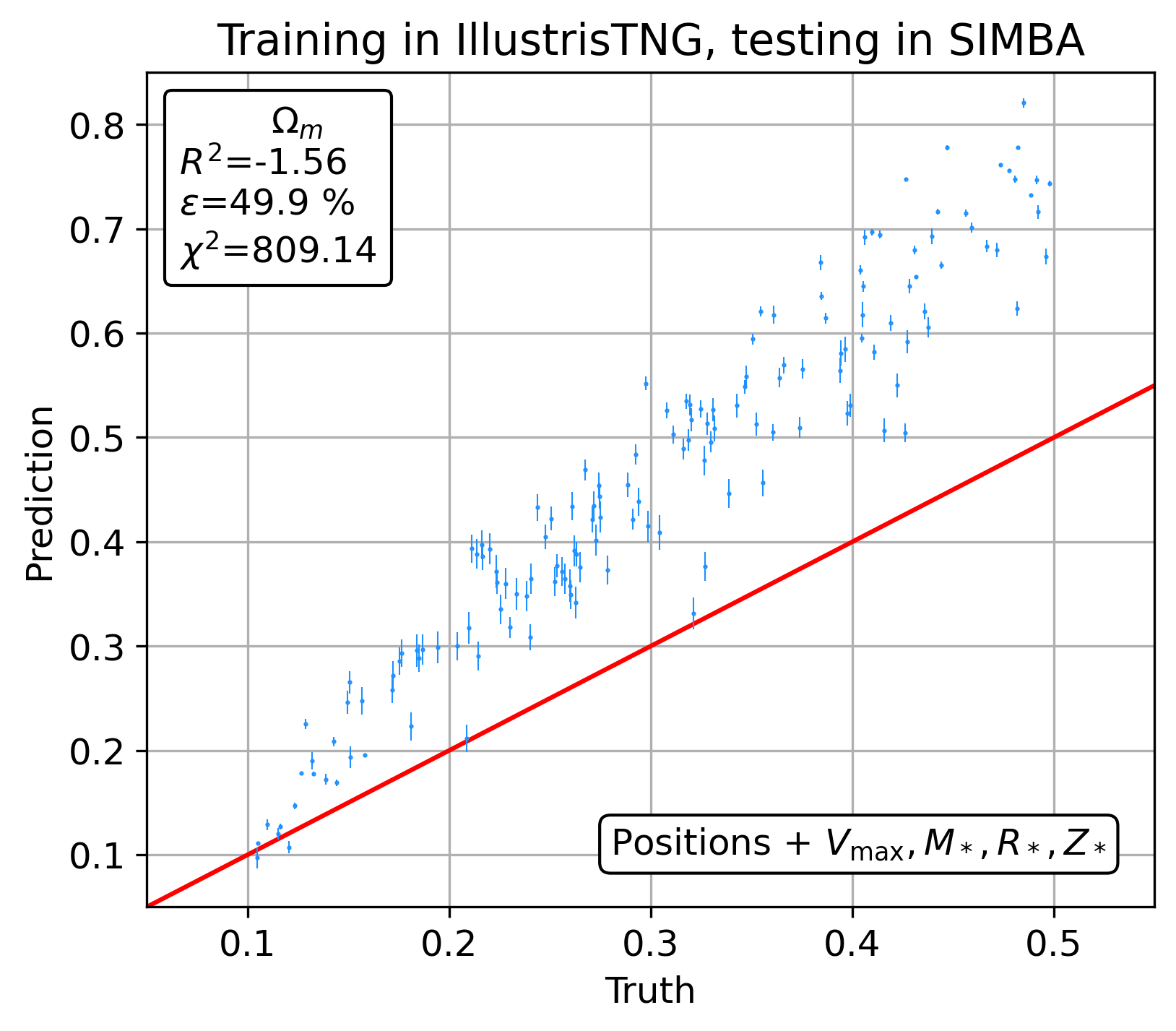}
\includegraphics[width=0.45\linewidth]{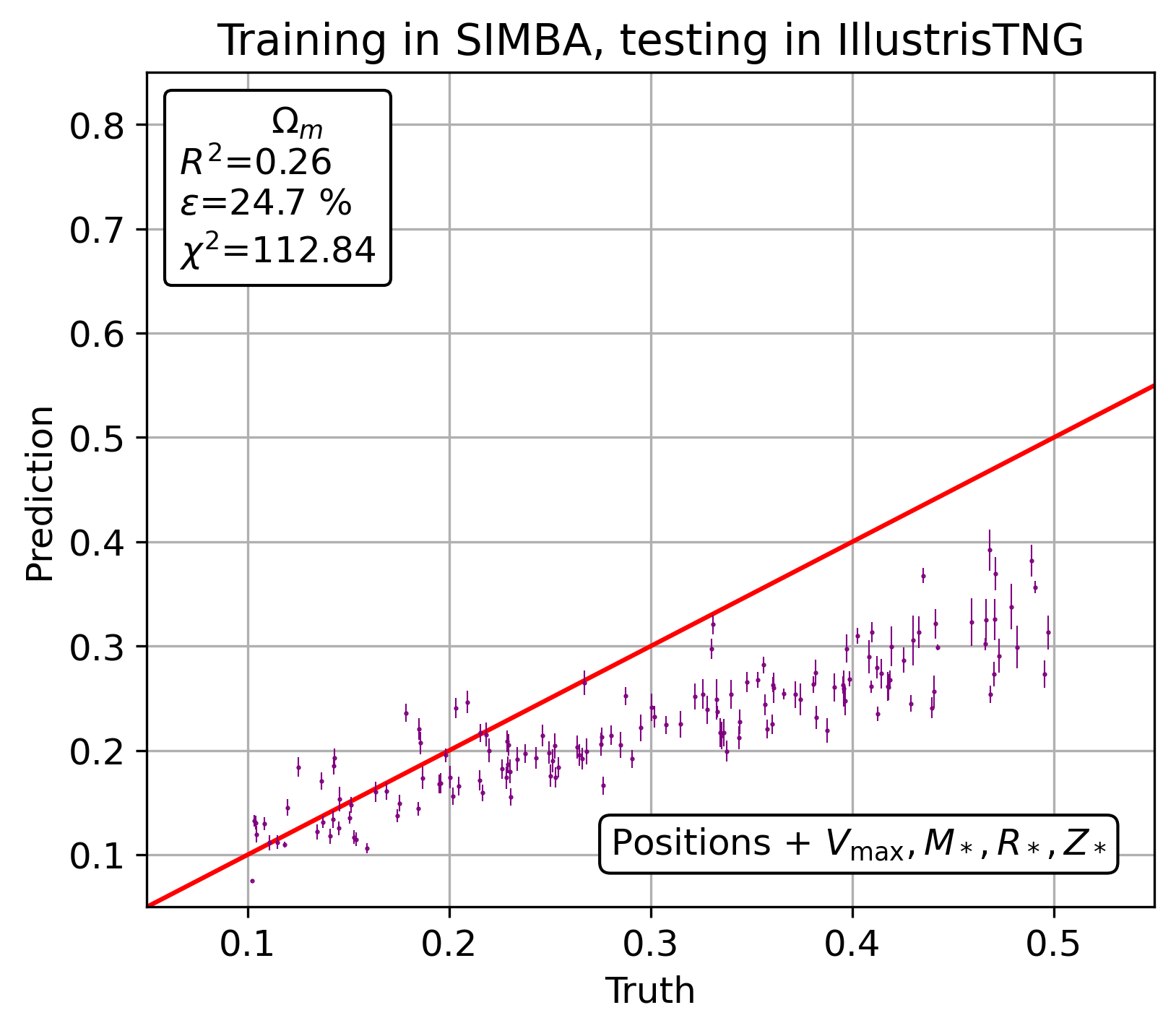}
\caption{Robustness test. We have tested the networks, trained on galaxy catalogues from the IllustrisTNG/SIMBA simulations, on catalogues from the SIMBA/IllustrisTNG simulations, showing the results in the left/right columns. As can be seen, our models are not robust in any of the considered scenarios. This can be due to the intrinsic differences that the distinct subgrid models imprint in both galaxy properties, abundance, and clustering. See text for details.}
\label{fig:crosstest}
\end{center}
\end{figure*}

The simulations in the two CAMELS suites, IllustrisTNG and SIMBA, use different subgrid physics models, which can be seen as two potential different approximations to reality. Our models are trained using a large variety of astrophysical models, so that the GNNs will learn to be insensitive to the particular baryonic effects of a given simulation. In other words, our models are trained to learn to marginalize over baryonic effects within an assumed subgrid physics model. While these ML algorithms learn to extract clustering and cosmological information within a given suite, it is not clear whether they could still perform well when the underlying subgrid physics model is different to the one used for training.

In order to check whether our models are indeed insensitive to the baryonic modeling, we can train a model on galaxy catalogues from IllustrisTNG simulations and test the model on catalogues from SIMBA simulations (and the other way around). We note that a model that is designed to be used with real data should pass this test at the level of simulations. This kind of tests are non trivial to succeed, and have been carried out in other previous works using CAMELS simulations with diverse levels of performance \citep{2021arXiv210910360V, 2022arXiv220102202V, GNN_CAMELS,2022arXiv220104142N, Helen_2021, Wadekar_2022}. 

We show in Fig. \ref{fig:crosstest} the results of the robustness test employing the models trained in Secs. \ref{sec:posonly} and \ref{sec:galacfeat}. Left column corresponds to networks trained in IllustrisTNG and cross tested in SIMBA, while right column corresponds to GNNs trained in SIMBA and tested in IllustrisTNG. We can see that the test fails in the two scenarios, i.e. for catalogues containing galaxy positions and for catalogues containing galaxy positions and properties. As can be seen in Fig. \ref{fig:crosstest}, relative errors raise up to $\sim 20$ \% when only positions (top row) or the combination $M_*$, $R_*$ and $Z_*$ (middle row) are employed, and can be even worse if $V_{\rm max}$ is also used (bottom row). The fact that the failure is less severe when only spatial distribution or the features $M_*$, $R_*$ and $Z_*$ are considered indicates that these properties are more robust than $V_{\rm max}$ regarding the baryonic effects. It is interesting to see that, while training in IllustrisTNG and testing in SIMBA leads to overpredictions, training in SIMBA and testing in IllustrisTNG produces the opposite result. 

It is perhaps not surprising that the models are not robust when using galaxy properties, as it is well known that the properties of the galaxies from the two different suites can be very different. \cite{2022arXiv220102202V} found that even at the level of a single galaxy properties in IllustrisTNG and SIMBA are different enough to make their model not robust. 

On the other hand, it is more difficult to explain why the model that only make use of galaxy positions is not robust. We believe that this may be due to the fact that our galaxy catalogues contain a large fraction of small galaxies with stellar masses below $10^9~h^{-1}M_\odot$ (see Fig. \ref{fig:jointplot}). The abundance and distribution of these galaxies may be very sensitive to the particular subgrid physics considered. For instance, the catalogues from SIMBA contain an average of 1200 galaxies, while this number lowers to 700 in the case of IllustrisTNG.

We thus conclude that our models are currently not robust enough, and therefore cannot be directly employed to perform parameter inference from galaxy catalogues of real data. The solution to this problem may not be simple and it is beyond the scope of this work.

%%%%%%%%%%%%%%%%%%%%%%%%%%%%%%%%%%%%%%%%%%%%%%%%%%%%%%%
%%%%%%%%%%%%%%%%%%%%%%%%%%%%%%%%%%%%%%%%%%%%%%%%%%%%%%%
\section{Conclusions}
\label{sec:discussion}

The spatial distribution of galaxies embeds precious information about fundamental physics. Unfortunately, we do not know which is the optimal procedure to extract that information, as galaxies follow, on small scales, a complex non-Gaussian pattern. On the other hand, neural networks can identify unique signatures that the cosmological parameters leave on those patterns, and therefore find optimal estimators to extract the maximum amount of information from the data. In this work we have used Graph Neural Networks (GNNs) to perform likelihood-free inference at the galaxy-field level while accounting for astrophysical uncertainties, as modelled in CAMELS.

We have argued that GNNs may be more appropriate for this task than other deep learning approaches such as Convolutional Neural Networks (CNNs) because 1) they are the natural architectures to deal with sparse and irregular data, like galaxy catalogues, 2) they can extract information from all scales, while CNNs are limited to scales larger than the pixel/voxel size, 3) it is easy to enforce the underlying symmetries such as translational\footnote{We note that translational invariance is also enforced with CNNs.} and rotational invariance.

The main take conclusions of this work are the following:

\begin{itemize}
\item We have shown that GNNs can be trained to predict non-trivial clustering properties of 3D distributions of points, like their power spectrum. This is our motivation to use GNNs to find optimal estimators that can extract the maximum information at the galaxy-field level.

\item We have trained GNNs to perform a field-level inference on the value of the cosmological parameters from galaxy catalogues generated from the hydrodynamic simulations of the CAMELS project while accounting for uncertainties arising from baryonic effects. We emphasize that our models are rotational and translational invariant, and can extract information from any available scale down to the minimum distance between galaxies: $\sim 5~h^{-1}$kpc in the CAMELS simulations at $z=0$.

\item We have shown that our models can infer the value of $\Omega_{\rm m}$ from galaxy catalogues covering volumes as small as $(25~h^{-1}{\rm Mpc})^3$ and containing only $\sim10^3$ galaxies with an accuracy between $\sim4\%$ to $\sim13\%$ at $z=0$. 

\item While our networks are able to extract information from the positions of galaxies alone, including other galaxy properties such as stellar mass, stellar radius, stellar metallicity, and maximum circular velocity can yield significantly tighter constraints. Most of the information our networks are extracting are due to galaxy properties and not to galaxy clustering.

\item Unfortunately, our models are not completely robust, i.e. a model trained on galaxy catalogues from IllustrisTNG simulations do not perform well when tested on galaxy catalogues from SIMBA simulations. Therefore, our models cannot be used with real data yet.
\end{itemize}

\subsection{Robustness}

This work highlights the large amount of cosmological information that resides on small, non-linear, scales, and how machine learning methods can be used to extract it. On the other hand, we also showed how the models do not extrapolate well, perhaps indicating that the models are extracting too much information. This is a big obstacle that prevents the deployment of our models to be used with real data. We note that the lack of robustness has been recently seen in other works involving the CAMELS simulations \citep[see e.g.][]{2021arXiv210909747V, 2022arXiv220102202V}. 

We emphasize that all our models exhibit a lack of robustness, although the models that only made use of galaxy positions are more robust than those employing all galactic features considered. There could be several reasons that explain the latter. First, even if the CAMELS simulations cover a very large volume in the astrophysical parameter space, perhaps the simulations in the IllustrisTNG and SIMBA suites do not fully overlap, so marginalizing over one does not guarantee the proper marginalization over the other. This can already be seen in Fig. \ref{fig:jointplot}, that shows how galaxy properties only overlap partially. Second, the networks may be learning numerical artifacts that are distinct in simulations with different subgrid physics. Third, perhaps the representation of the data is slightly different in the two simulations, leading to different dimensions where simulations do not overlap.

The fact that the models trained using only galaxy positions are also not robust is more worrisome. This may be happening because our galaxy catalogues contain a non-negligible fraction of small galaxies, whose abundance and clustering may be largely affected by baryonic effects. If this would be the case, these small galaxies can be seen as a challenge for hydrodynamic simulations. We believe it is important to investigate whether our conclusions hold when using galaxy catalogues containing more massive galaxies. 

The volume covered by the galaxy catalogues may also be an important piece affecting the model robustness. On sufficiently large scales, we would expect the models using only galaxy positions to be robust. However, in this case, the field-level inference may not bring substantial improvement over the traditional power spectrum. If the problem is instead coming from the clustering on small scales, it will be important to develop models that do not make use of that information. For instance, one can create graphs from galaxy catalogues by setting edges only if the distance between two galaxies is above some scale.

\subsection{Future work}

We outline some tasks to be carried out in future work:
\begin{itemize}

\item In this work we have worked with galaxy catalogues in real-space. However, galaxies are observed in redshift-space. It will be interesting to quantify how much information is gained, or lost, by performing the analysis with galaxy catalogues in redshift-space. In this case the symmetries of our models should be changed: the axis along which redshift-space distortions is implemented will break down the rotational symmetry considered here.

\item We have made use of galaxy catalogues at $z=0$ for inferring $\Omega_m$. It is important to train networks with galaxy catalogues at redshifts higher than 0 and investigate how much information is gained, or lost, and whether the models are more robust at higher redshifts.

\item It is critical to quantify the effects of volume in our results. Using galaxy catalogues covering larger volumes will help addressing 1) how much information is coming from clustering versus internal galaxy properties as function of scale, 2) whether the models are more robust with larger volumes or more massive galaxies, and 3) whether the model can infer $\sigma_8$ and potentially other parameters. Galaxy catalogues from CAMELS-SAM \citep{2022arXiv220402408P} and Molino \citep{Molino}, among others, will be perfect for this task.

\item In this work we have not taken into account the effects of super-sample covariance, i.e., the effect of perturbations on scales larger than the volume of our simulation boxes. Addressing how much information is gained or lost by this is not only interesting, but needed in order to build models that can be used with real data. This can be achieved in several ways, like running simulations with larger volumes or using separate-Universe simulations \citep{Li_2014, Li_2018, Barreira_2019}.

\item It will also be interesting to develop more interpretable models. For instance, for GNNs with one, or few, blocks, it may be possible to use symbolic regression to find analytic expressions that approximate the behaviour of the network \citep{Cranmer:2020wew, Lesmos_2022}. In this case, it may be possible to understand what operation the network is doing and make it robust to changes in subgrid physics.

\item It would be important to investigate whether training in galaxy catalogues from both IllustrisTNG and SIMBA make the models more robust, and whether they will work on simulations from a third suite. Alternatively, contrastive learning techniques can be applied to build models which are by construction insensitive to the differences between subgrid models.

\end{itemize}

\section*{Data Availability}

The implementation of the GNNs underlying this article, \textsc{CosmoGraphNet} \citep{pablo_villanueva_domingo_2022_6485804}, is available on \href{https://github.com/PabloVD/CosmoGraphNet}{GitHub \faGithub} \footnote{\url{https://github.com/PabloVD/CosmoGraphNet}}. Details on the CAMELS simulations can be found in \url{https://www.camel-simulations.org}.

\section*{Acknowledgements}

We thank the CAMELS team for useful conversations. The work of FVN is supported by the Simons Foundation. The training of the GNNs has been carried out using GPUs from the Tiger cluster at Princeton University.

\appendix

\section{Power spectrum extrapolation}
\label{sec:pstest}

\begin{figure*}[th!]
\begin{center}
\includegraphics[width=0.4\linewidth]{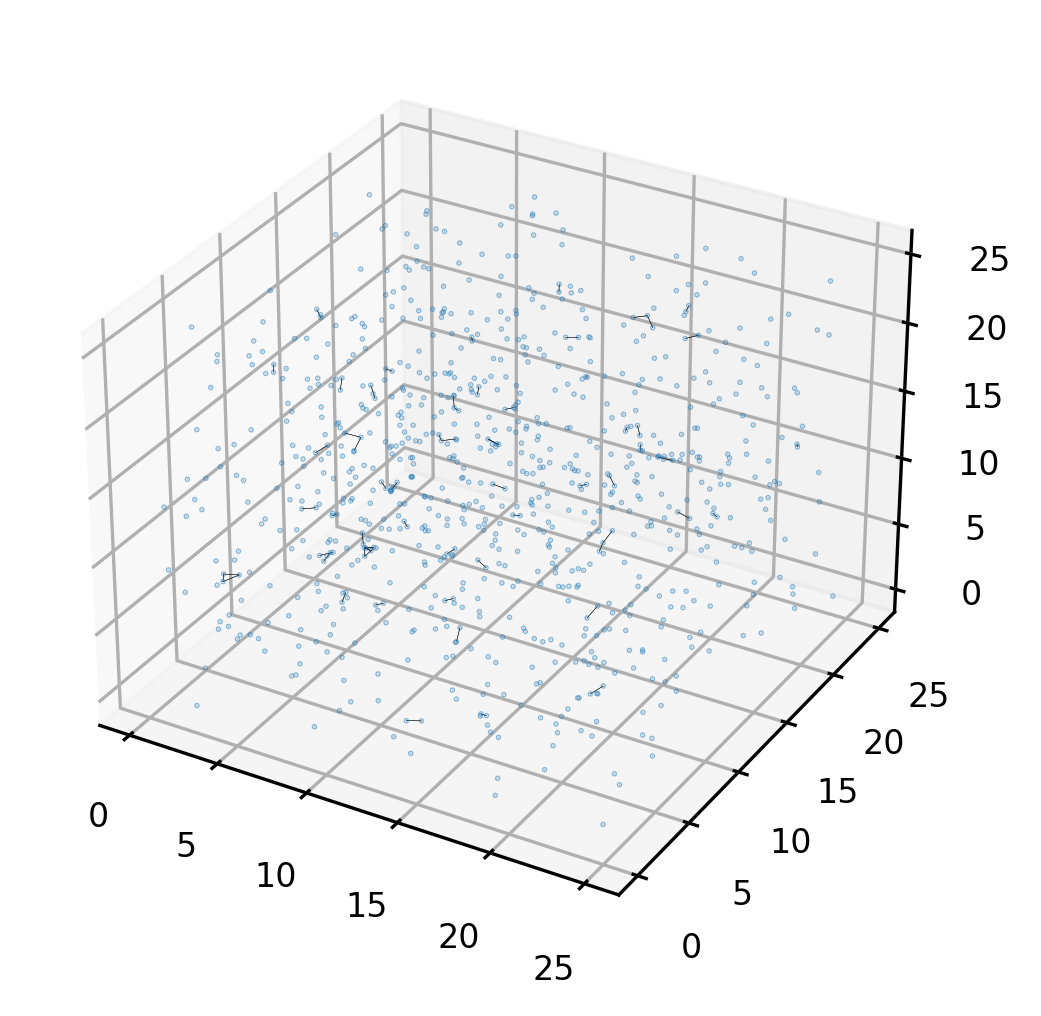}
\includegraphics[width=0.4\linewidth]{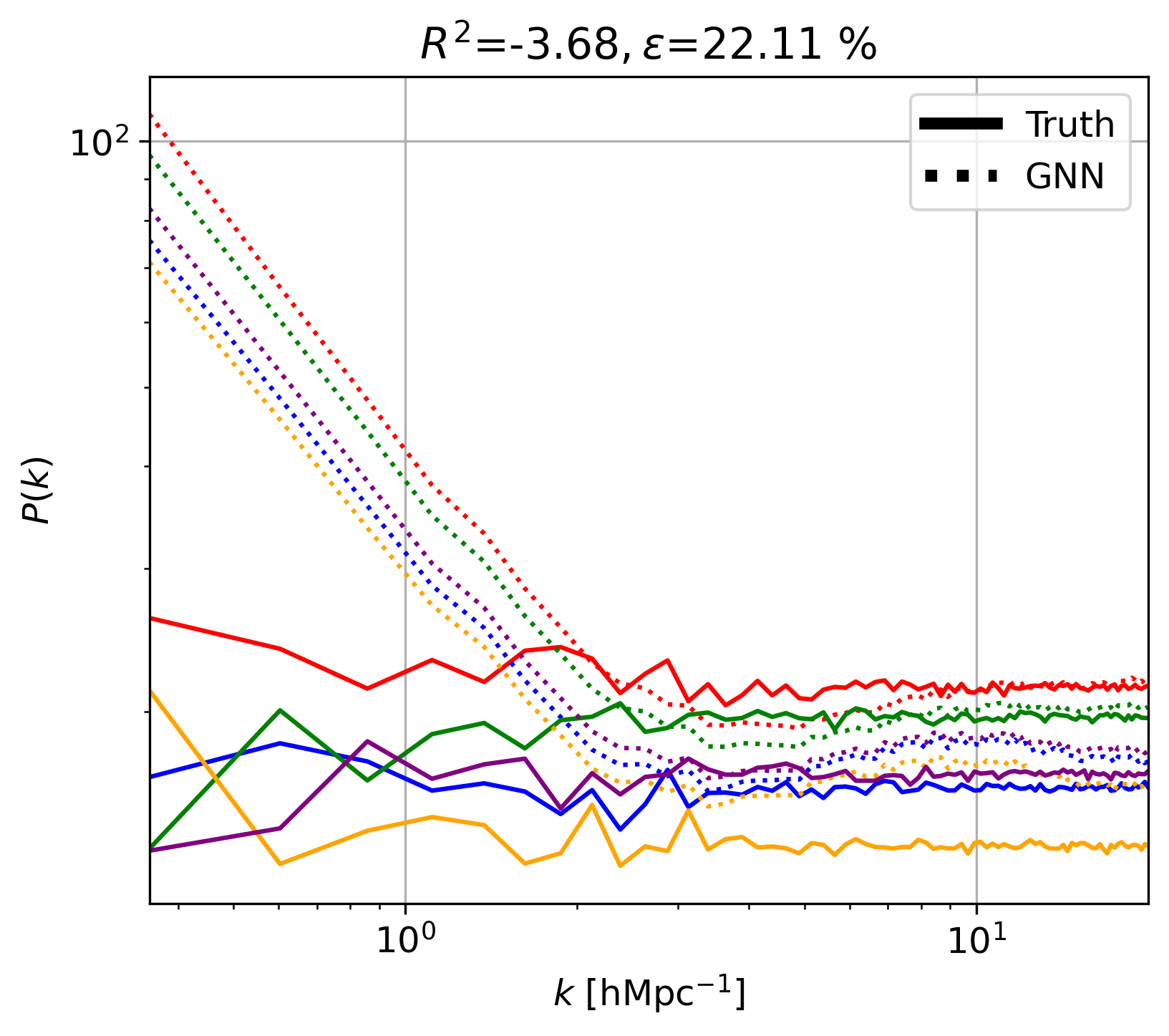}
\includegraphics[width=0.4\linewidth]{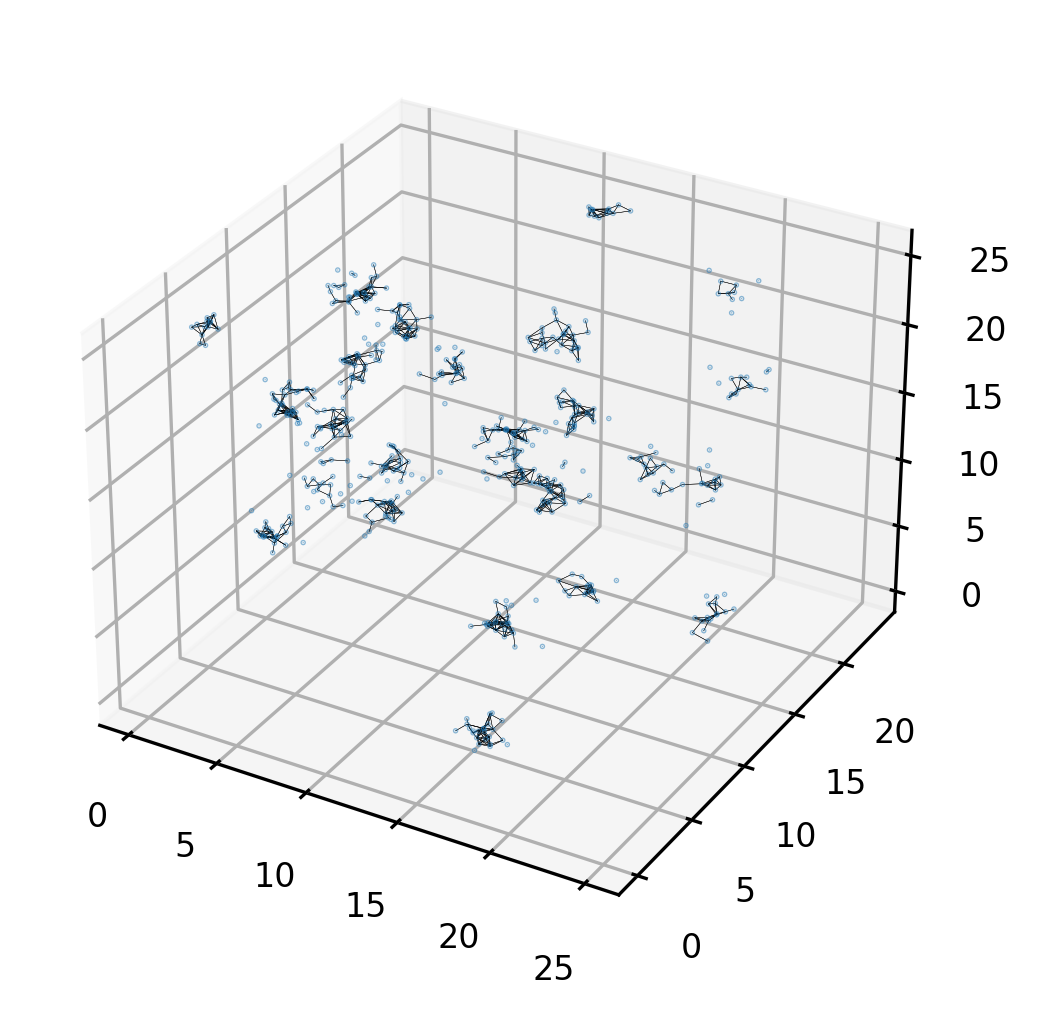}
\includegraphics[width=0.4\linewidth]{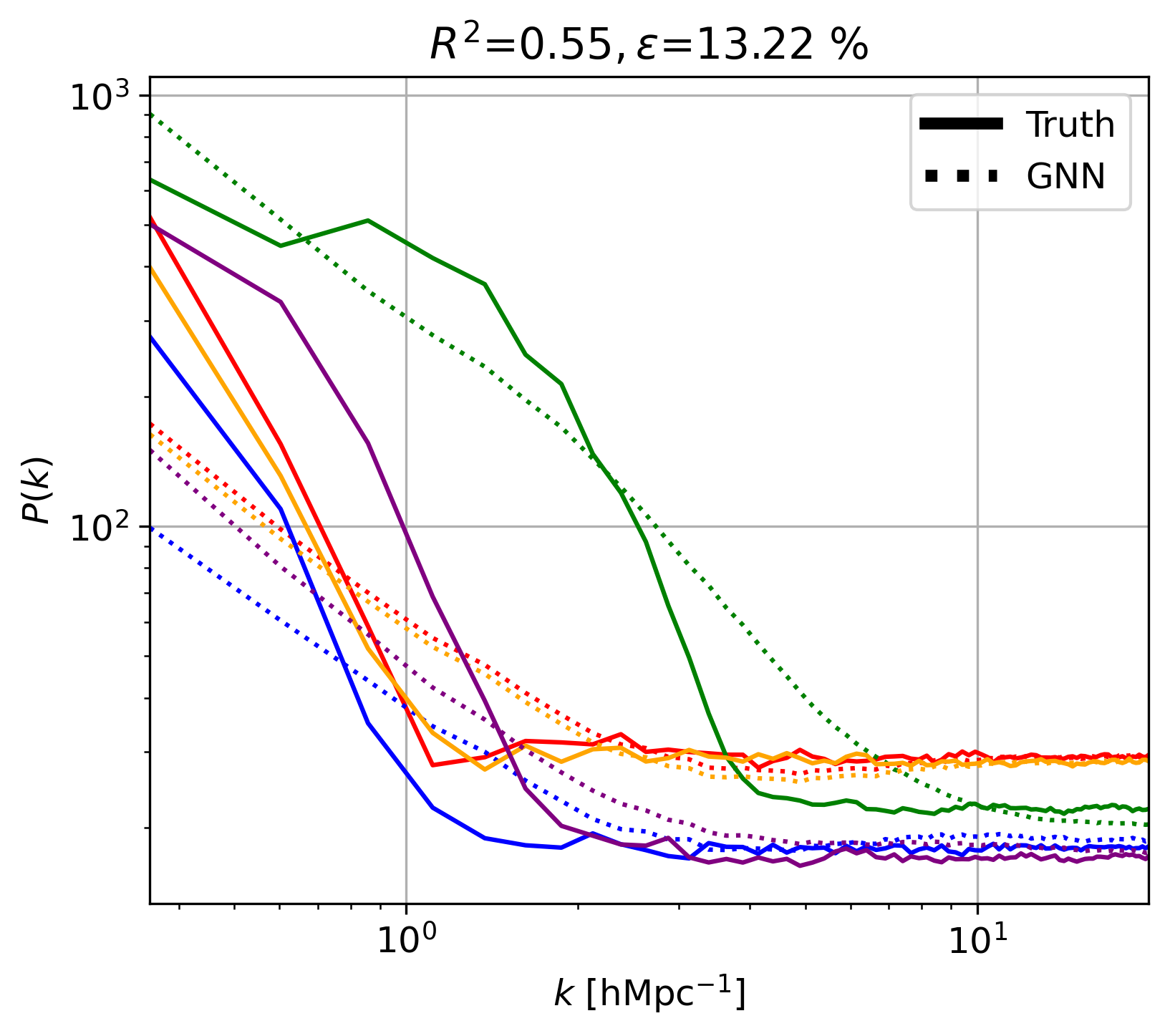}
\includegraphics[width=0.4\linewidth]{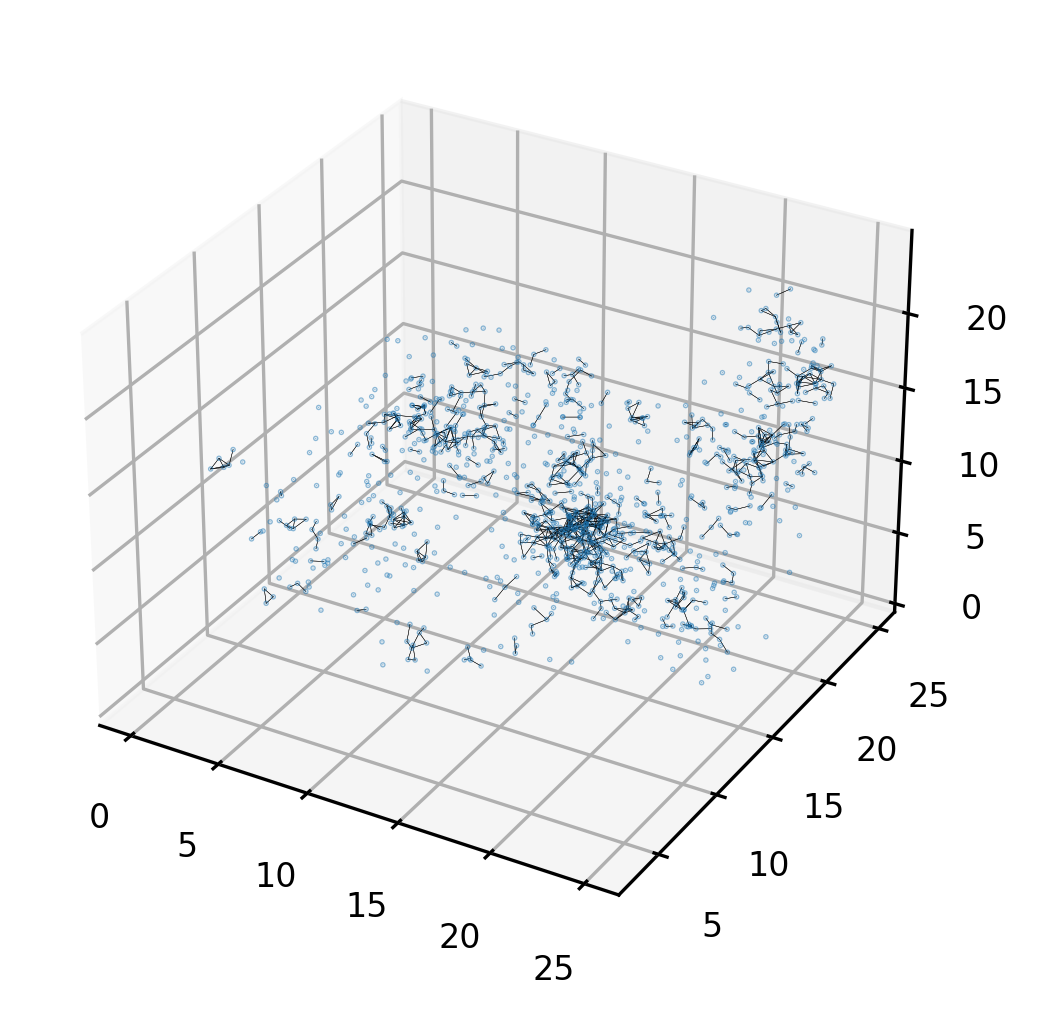}
\includegraphics[width=0.4\linewidth]{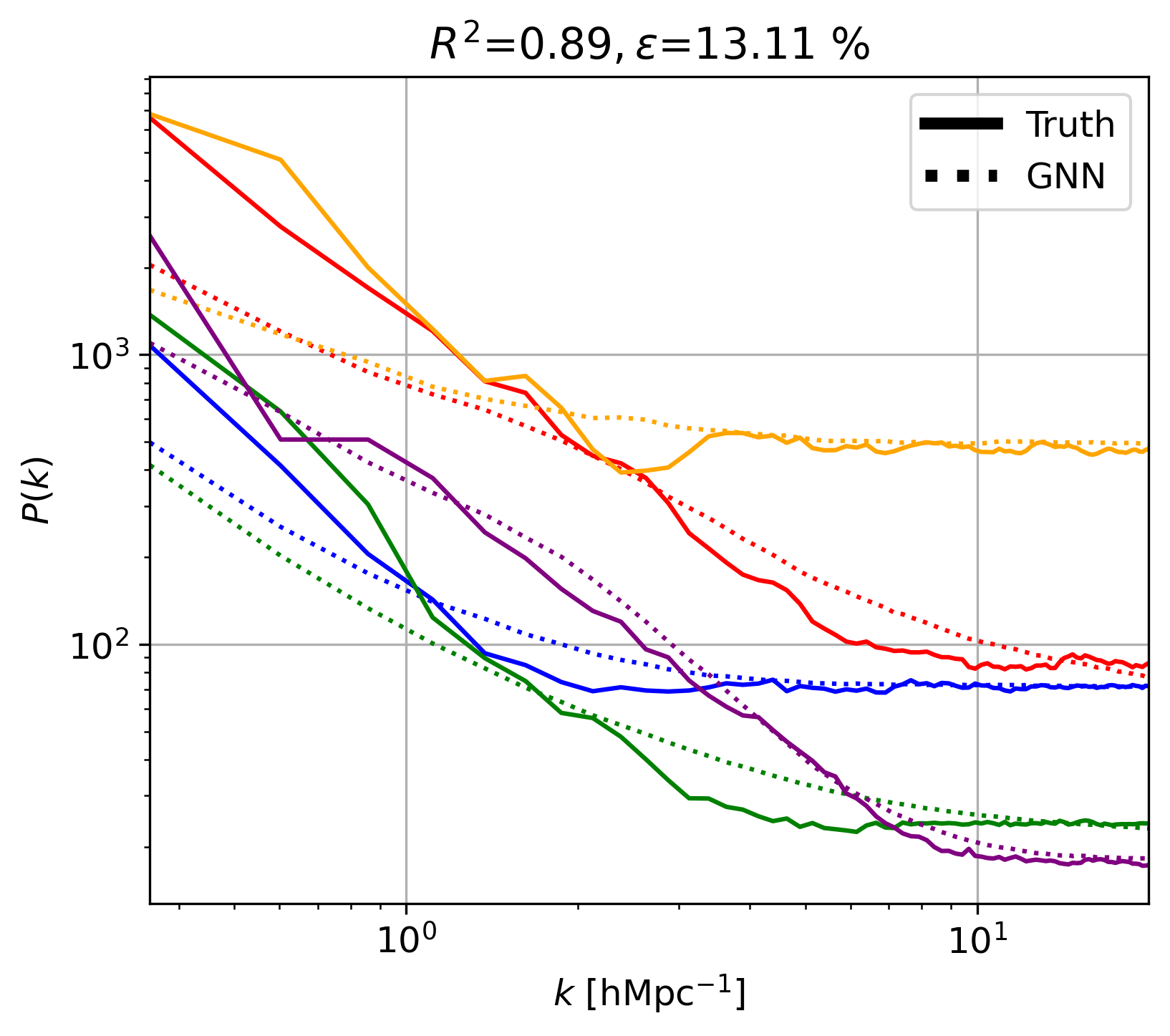}
\caption{We have quantified how accurate our GNN, that has been trained on galaxy catalogues to predict its power spectrum, extrapolates by using it to predict the power spectrum of point distributions with very different clustering properties: Poisson process (top), Neyman-Scott (middle), and Soneira-Peebles fractal hierarchical clustering (bottom). The left column shows examples of the graphs for these point distributions, while the right column displays the true power spectra (solid) versus the prediction of the model (dotted). As can be seen, the model is not very accurate at predicting the correct amplitude of the power spectrum on large scales. We believe that this may happen because the model has learned some correlations that are only present in the galaxy catalogues it was trained on. See text for details.}
\label{fig:ps_test}
\end{center}
\end{figure*}

In Sec. \ref{sec:ps} we trained a GNN using CAMELS galaxy catalogues to predict their power spectrum. We now investigate whether our model has really learned to compute the power spectrum by testing it on distributions of points with very different clustering patterns to the ones used for training. For this task we consider catalogues with three different clustering models:

\begin{itemize}

    \item \textbf{Poisson process}. This is simply a random distribution of points within the considered volume. It is well known that such distribution exhibits a white-noise power spectrum with an amplitude equal to $V/N$, where $V$ is the volume and $N$ the number of points.
    
    \item \textbf{Neyman-Scott process} \citep{1952ApJ...116..144N}. This is a clustered distribution formed by two nested Poisson processes. A set of cluster centers are randomly spread within a box following a Poisson process. Then, a cluster of points is distributed around each center, following a given distribution, that we take as a Gaussian in this case. The correlation function can be computed analytically, resulting in a Gaussian function \citep{Illian2008StatisticalAA}. The power spectrum can thus be derived from the correlation function, showing a Gaussian dependence on the wavenumber $k$.
    
    \item \textbf{Soneira-Peebles model} \citep{1977ApJ...211....1S, 1978AJ.....83..845S, 1980lssu.book.....P}. This is a fractal hierarchical distribution and is built as follows. A set of cluster centers are first randomly placed. Then, $\eta$ daughter points are placed around each cluster center following a Gaussian distribution, with a width given by a radius $R$. Each of these points becomes the center of a new cluster, around which $\eta$ new points are placed within a width $R/\lambda$. The process is repeated for $L$ times, generating a multilevel structure of clusters where at a level $l$ there is a cluster with size $R/\lambda^l$. Given its self-similar nature, this point process leads to a scale-free power spectrum (a power law in $k$) in the range $1/R<k<1/(R/\lambda^L)$.
\end{itemize}

For each clustering model we generate 50 catalogues, each of them having between 700 and 1200 points, similar to the mean number density of galaxies in the CAMELS catalogues. We then feed these catalogues to the GNN, that has been trained in CAMELS galaxy catalogues, to obtain their power spectrum. We compute the true power spectrum of these point catalogues using \textsc{Pylians} and we show in Fig. \ref{fig:ps_test} the comparison between the true power spectra and the network predictions. We can see that the network is partially able to predict the amplitude of the power spectrum on small scales, $k>5-10~h{\rm Mpc}^{-1}$. Even though a small offset can typically appear with respect to the true power spectrum in the Poisson scenarios, it works well for the  Neyman-Scott and Soneira-Peebles catalogues.

On large scales, however, the GNN is barely able to recover the true power spectrum. There are several reasons that may explain why the network fails in this regime (at low $k$). We may be tempted to justify this behaviour due to the presence of cosmic variance. However, we can see that the predicted shape of the power spectrum is pretty similar in all cases, being different to the true one. We suspect that since the model has been trained on galaxy catalogues from hydrodynamic simulations, were the volume is small and all scales may be tightly coupled, the model may be using these correlations to determine the amplitude and shape of the power spectrum on large scales. It may be possible that the model is just using information from small scales to determine the amplitude of clustering on those scales and then made use of the correlations of small and large scales to infer the power spectrum on large scales.

While in Sec. \ref{sec:ps} it has been shown how GNNs can learn to accurately compute non-trivial summary statistics from a distribution of points, the tests above illustrates how the learned procedure may not truly correspond to the traditional methods, and therefore, fails when extrapolating it. This could happen because the network is making use of information that is characteristic of the data it is trained on. If those features are not present in other datasets the model may fail. For instance, the shot noise spectrum is not well recovered in some cases. While a GNN trained on these kind of catalogs could trivially predict that regime (since it is basically given by the number of galaxies), the shot noise contribution in the training CAMELS catalogs is negligible (see Fig. \ref{fig:powerspec}). Training the model on data with many different clustering properties on different scales can force the network to learn more robust and close approximations to the true power spectrum estimator. 

We emphasize that the purpose of this work is not to show that GNNs can learn to compute the power spectrum for generic point distributions, but that in the case of galaxy catalogues from hydrodynamic simulations, they can learn to compute non-trivial clustering properties that may be sensitive to the value of the cosmological parameters. Developing a generic power spectrum estimator would involve training in point catalogues with very different clustering properties, number of points, volumes and other considerations, and it is out of the scope of this work.

\section{Constraints on $\sigma_8$}
\label{sec:sigma_8}

\begin{figure*}%[b!]
\begin{center}
\includegraphics[width=0.49\linewidth]{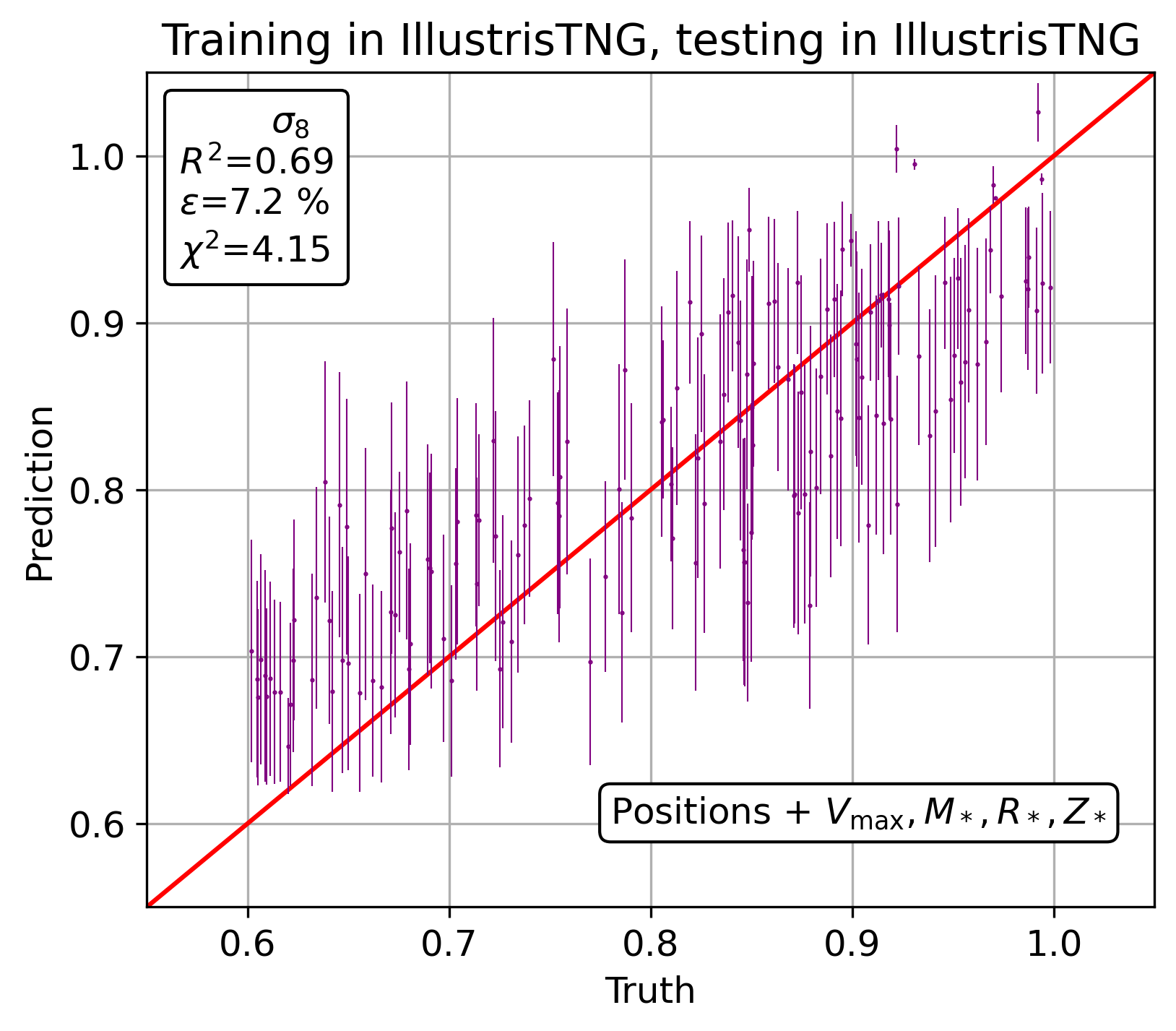}
\includegraphics[width=0.49\linewidth]{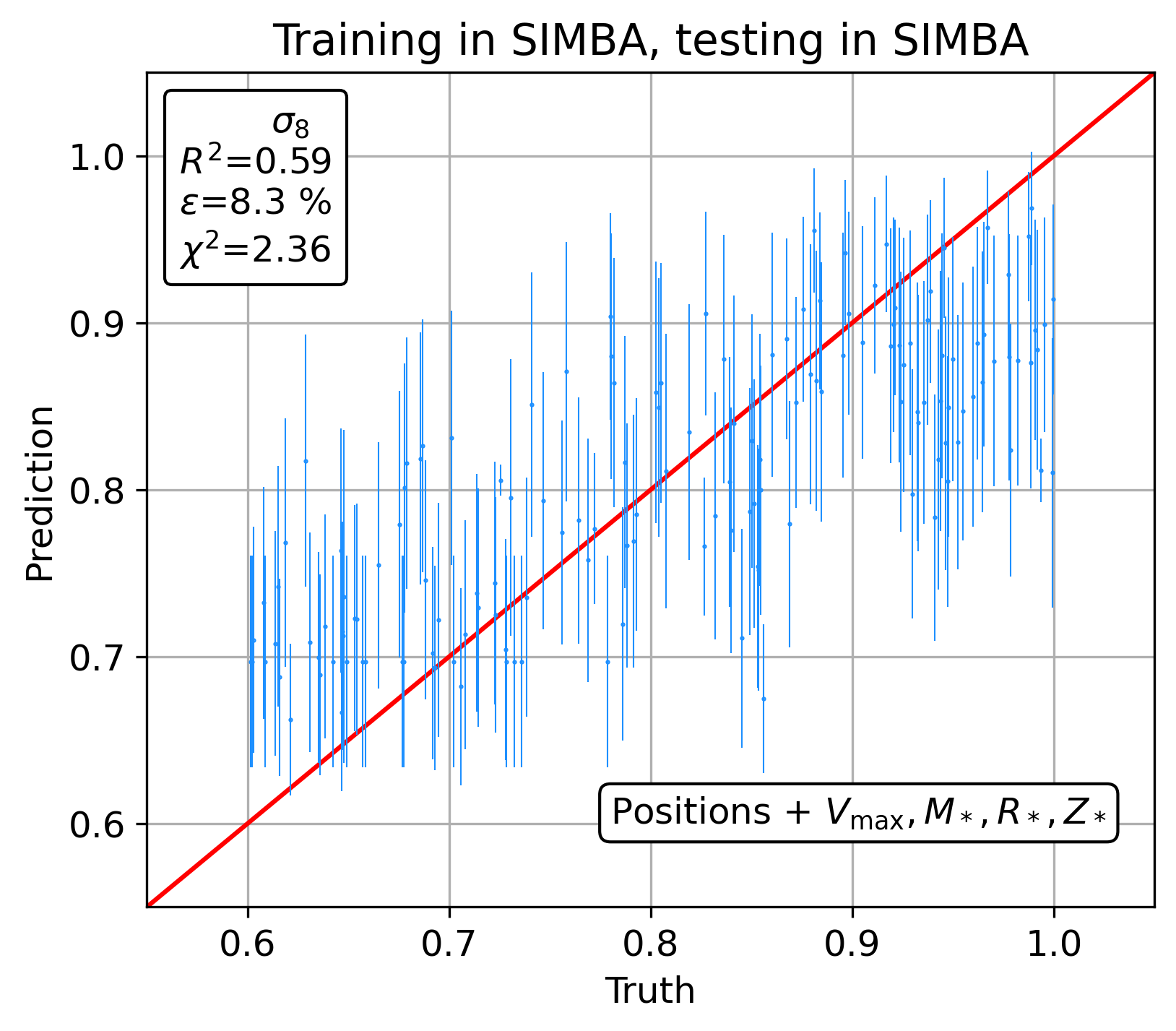}
\caption{We have trained GNNs to infer the value of $\sigma_8$ from galaxy catalogues that include galaxy positions, $M_*$, $R_*$, $Z_*$ and $V_{\rm max}$ on simulations from the IllustrisTNG (left) and SIMBA (right) suites. As can be seen, our models are not able to infer the value of $\sigma_8$ accurately.}
\label{fig:sigma8}
\end{center}
\end{figure*}

Here we report the results we obtain when training GNNs to infer the value of $\sigma_8$ using galaxy catalogues that contain both galaxy positions and galaxy properties ($M_*$, $R_*$, $Z_*$, and $V_{\rm max}$). We show the results in Fig. \ref{fig:sigma8} for the IllustrisTNG (left) and SIMBA (right) simulations. As can be seen, our models are unable to provide accurate constraints on $\sigma_8$. There are several reasons that could explain such bad performance. On the one hand, this could be due to degeneracies with other parameters. For instance, the cosmological effects resulting from varying $\sigma_8$ could be mimicked by modifying one or some of the astrophysical parameters considered in CAMELS, and in such case, the network would not be able to discern the specific value. On the other hand, the volume of the simulation boxes, $(25~h^{-1}{\rm Mpc})^3$, may be to small to properly capture the effect of $\sigma_8$ when the field is sampled with galaxies. Training our models on simulations covering larger volumes should enable us to constrain $\sigma_8$ with higher accuracy. We will explore this in future work. 

We have checked that training only with galaxy positions leads to even worse results, where the network hardly predicts outputs other than a constant mean value with large error bars. It is worth to note that $\sigma_8$ has been predicted with far better accuracy from different 2D fields in CAMELS \citep{2021arXiv210910360V, 2021arXiv210909747V}, while \cite{2022arXiv220102202V} showed worse performance than ours using properties from individual galaxies.

\bibliography{bibliography}

\end{document}